%% file: main.tex
\documentclass[sigconf]{acmart}

\usepackage{booktabs} 
\usepackage{multirow}
\usepackage{algorithmic}
\input{mmstyle}

\usepackage{makecell}
\usepackage{makecell}
\usepackage{xcolor}
\usepackage[ruled]{algorithm2e} 

\input{macros.tex}

\AtBeginDocument{%
	\providecommand\BibTeX{{%
			\normalfont B\kern-0.5em{\scshape i\kern-0.25em b}\kern-0.8em\TeX}}}

\setcopyright{acmcopyright}
\copyrightyear{2023}
\acmYear{2023}
\acmConference[Los Angeles '23]{Los Angeles '23: ACM SIGGRAPH}{August 06--10, 2023}{Los Angeles, USA}
\acmDOI{10.1145/3588028.3603647}
\acmISBN{9798400701528}
\acmBooktitle{Special Interest Group on Computer Graphics and Interactive Techniques Proceedings (SIGGRAPH '23 Proceedings), August 06--10, 2023, Los Angeles, CA, USA}
\begin{document}
\title{Dynamic Storyboard Generation in an Engine-based Virtual Environment for Video Production}

\author{Anyi Rao}\authornotemark[1]
\affiliation{%
	\institution{Stanford University}
	\country{USA}
}
\email{anyirao@stanford.edu}

\author{Xuekun Jiang} \authornotemark[1]
\affiliation{%
	\institution{Shanghai AI Lab}
	\country{China}
}
\email{jiangxuekun@pjlab.org.cn}

\author{Yuwei Guo} 
\affiliation{%
	\institution{Shanghai AI Lab}
	\country{China}
}
\email{guoyuwei@pjlab.org.cn}

\author{Linning Xu} 
\affiliation{%
	\institution{CUHK}
	\country{Hong Kong, China}
}
\email{linningxu@ie.cuhk.edu.hk}

\author{Lei Yang} 
\affiliation{%
	\institution{Shanghai AI Lab}
	\country{China}
}
\email{yanglei@pjlab.org.cn}

\author{Libiao Jin} 
\affiliation{%
	\institution{CUC}
	\country{China}
}
\email{libiao@cuc.edu.cn}

\author{Dahua Lin} 
\affiliation{%
	\institution{Shanghai AI Lab, CUHK}
	\country{Hong Kong, China}
}
\email{dhlin@ie.cuhk.edu.hk}

\author{Bo Dai}
\affiliation{%
	\institution{Shanghai AI Lab}
	\country{China}
}
\email{daibo@pjlab.org.cn}


\input{articles/abstract.tex}

\maketitle


\input{articles/introduction.tex}
\input{articles/related.tex}
\input{articles/method.tex}
\input{articles/experiment.tex}
\input{articles/conclusion.tex}
\bibliographystyle{ACM-Reference-Format}
\bibliography{main}

\clearpage

\end{document}

%% file: mmstyle.tex
\usepackage{amsmath}
\usepackage{xspace}

\providecommand{\eg}{\textit{e.g.}\@\xspace}
\providecommand{\ie}{\textit{i.e.}\@\xspace}

\providecommand{\etal}{\textit{et al}\@\xspace}

\newcommand{\va}{\mathbf{a}}

\newcommand{\vc}{\mathbf{c}}

\newcommand{\vp}{\mathbf{p}}

\newcommand{\cC}{\mathcal{C}}

\newcommand{\cL}{\mathcal{L}}

\newcommand{\cS}{\mathcal{S}}


%% file: macros.tex
\usepackage{xcolor}
\usepackage{wrapfig}
\usepackage{soul}

\definecolor{yellow}{rgb}{1,1, 0.6}
\definecolor{lightyellow}{rgb}{1,1, 0.8}
\definecolor{orange}{rgb}{1, 0.8, 0.6}
\definecolor{coral}{RGB}{246,131,65}
\definecolor{pinkred}{rgb}{1, 0.6, 0.6}
\definecolor{hotpink}{RGB}{238,64,195}
\definecolor{amber}{rgb}{1.0, 0.49, 0.0}
\definecolor{lavender}{RGB}{207,226,243}
\definecolor{gainsboro}{RGB}{208,224,227}
\definecolor{gainsboro2}{RGB}{217,234,211}
\definecolor{blanchedalmond}{RGB}{252,229,205}

\newcommand{\model}{VDS}
\newcommand{\modelss}{VDS\xspace}

%% file: articles/abstract.tex
\ccsdesc[100]{Information systems~Multimedia content creation}

\keywords{virtual storyboard, cinematography, neural networks}

\begin{teaserfigure}
	\vspace{-3pt}
	\begin{center}
		\includegraphics[width=\linewidth]{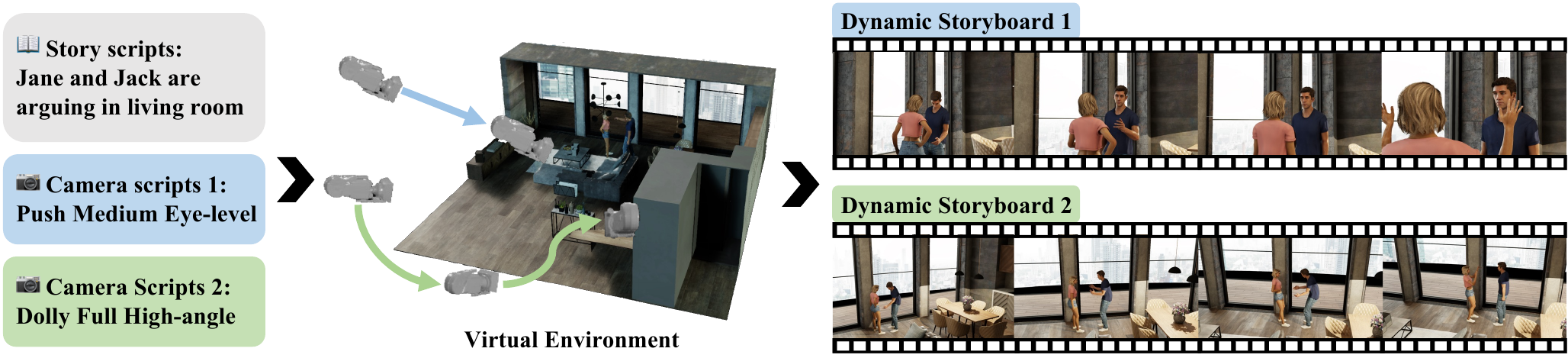}
	\end{center}
	\vspace{-13pt}
	\caption{We present Virtual Dynamic Storyboard (VDS)
		that takes user input \emph{story} and \emph{camera} scripts and automatically composes dynamic storyboards in an engine-based virtual environment for pre-visualization. 
		Here we show two results produced by \modelss with top-ranked scores of quality. 
		Video demos can be found in the supplementary.
	}
	\label{fig:teaser}
	\vspace{3pt}
\end{teaserfigure}

\begin{abstract}
	\label{sec:abs}
	Amateurs working on mini-films and short-form videos usually spend lots of time and effort on the multi-round complicated process of setting and adjusting scenes, plots, and cameras to deliver satisfying video shots.
	We present Virtual Dynamic Storyboard (\model) to allow users storyboarding shots in virtual environments, 
	where the filming staff can easily test the settings of shots before the actual filming.
	\modelss runs on a ``propose-simulate-discriminate'' mode:
	Given a formatted story script and a camera script as input, 
	it generates several character animation and camera movement proposals following predefined story and cinematic rules
	to allow an off-the-shelf simulation engine to render videos.
	To pick up the top-quality dynamic storyboard from the candidates, 
	we equip it with a shot ranking discriminator based on shot quality criteria learned from professional manual-created data. 
	\modelss is comprehensively validated via extensive experiments and
	user studies, demonstrating its efficiency, effectiveness, and great potential in assisting amateur video production.
\end{abstract}

%% file: articles/introduction.tex
\section{Introduction}
\label{sec:intro}

\begin{figure*}[!t]
	\centering
	\includegraphics[width=\linewidth]{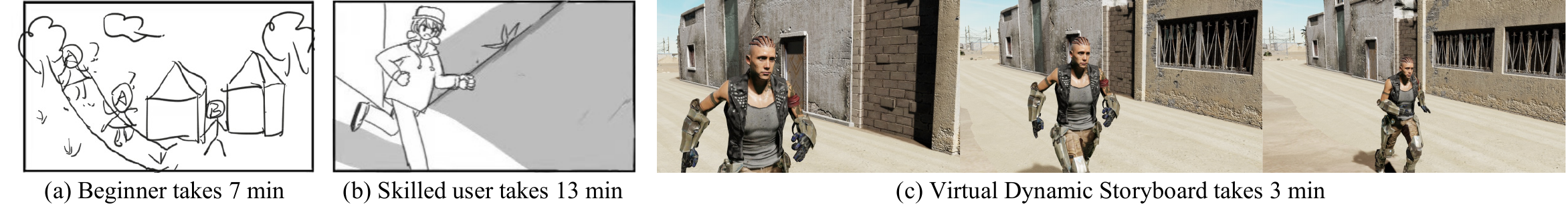}
	\vspace{-25pt}
	\caption{(a) Traditional hand-paint storyboards require time efforts and drawing skills; (b) our virtual dynamic storyboard can help amateurs to acquire desired results efficiently.}
	\vspace{-11pt}
	\label{fig:storyboard}
\end{figure*}

Born in 1930, storyboarding technique helps video directors and cinematographers design each shot\footnote{Video is usually made up of several shots, where each one is a series of visual continuous frames.}, 
figure out potential problems, 
and communicate ideas to save time and resources raised in practical video production~\cite{storyboard}.
A conventional storyboard (Fig.~\ref{fig:storyboard}(a,b)) 
represents each shot with one or two still frames,
depicting the scene layout, character actions, as well as camera parameters such as scale and angle.
However,
such a static storyboard is often dry and rigid,
since it lacks the intrinsic capability to faithfully demonstrate dynamic semantics including character and camera movements.
Although arrows and textual instructions can be used to indicate movement directions as amendments,
they often lead to significant semantic ambiguity in the final static storyboard.
Moreover,
most conventional storyboards are hand-painted sketches that require both time and drawing skills.
A rudimentary storyboard of a target video lasting for minutes can take artists hours to finish,
let alone amateur creators who lack skills.

Given the aforementioned drawbacks,
it is thus of great demand to improve conventional storyboard generation from two aspects:
1) producing dynamic storyboards instead of static ones, 
where the demonstration of dynamic semantics is straightforward.
2) building a handy tool that can create customized storyboards in a semi-auto way.
There are two candidate solutions, namely neural video generative models as well as virtual cinematography.
While neural video generation models have made significant progress in recent years \cite{singer2022make,he2022latent},
their outputs still suffer from frame-wise spatial-temporal inconsistency,
let alone meet the cinematic requirements of dynamic storyboards.
A better alternative is virtual cinematography \cite{he1996virtual},
which includes three modules, 
real-time executor to drive characters move, 
virtual cinematographer to handle the visual layout, 
and render to output the results,
which can be adapted for storyboard generation.
Still, two significant problems need to be further addressed in order to fulfill this task with enough high quality.
First, most existing works~\cite{shah2018airsim, fabbri2021motsynth,jiang2021example,jiang2020example} 
only study the camera behavior under \emph{fixed} plots and environments that require heavy manual force to modify.
Secondly, the full action space of camera movements is too large to allow effective per-frame decisions~\cite{truong2018extracting}. 
The camera action space in filming consists of seven degrees of freedom (7DoF), 
where a decision on camera position, rotation, and focal length should be determined for each frame.
Moreover, abrupt changes can easily occur when creating video frame by frame, 
yet one of our goals is to produce smooth videos with harmonious content and cinematic styles.

In this paper,
we propose a novel virtual dynamic storyboard approach in 
an efficient and customized way for amateurs, as shown in Fig.\ref{fig:pipeline}.
It follows a ``propose-simulate-discriminate'' paradigm.
For each shot,
it first translates the inputted high-level \emph{story and camera script} in the form of $(\langle$char$\rangle$ $\langle$do$\rangle$ $\langle$sth/swh$\rangle)$ 
tuples for story scripts and $(\langle$movement$\rangle$ $\langle$scale$\rangle$ $ \langle$angle$\rangle)$
tuples for camera scripts,
into proposals of shot hyperparameters.
Each proposal is then executed by the simulation module built on top of a modern graphic engine (\eg, Unity, Unreal, Omniverse~\cite{unity, unreal, omniverse}) 
to acquire its corresponding rendered output.
Finally,
a data-driven discriminator is adopted to rank all these rendered outputs 
as a recommendation for users to choose the final dynamic demonstration of this shot. 
Our key insights are as follows:
1) The enormous action space for all possible character and camera trajectories on a frame-by-frame basis 
can easily lead to low-quality shot candidates in terms of inter-frame abruptness 
and less meaningful camera motions, let alone the desired artistic feelings.
Our shot-based story and camera proposal generation significantly reduces the search space for appropriate camera configurations, further ensuring the results are more plausible.
2) By wrapping the graphic engine into a simulation module,
the proposed approach significantly improves its accessibility and computational efficiency,
thanks to engines' rich and highly-structured functionalities and highly-optimized computation pipeline.
Moreover,
the proposed approach also readily enjoys a series of additional benefits brought by the graphic engine,
such as function-wise scalability and high rendering quality.
3) There might be several plausible shots satisfying the requirements of story and camera scripts. 
No universal criteria can serve as direct supervision and evaluation metrics for generated shots.
We resort to a data-driven learning approach that aims to automatically discover rules from professional manual-created clips.
with a \emph{shot ranking discriminator}.
It learns from a large amount of data to evaluate shots by distinguishing between different quality samples.
Through the \emph{class-aware contrastive objectives}, our approach can successfully select shots with top-ranked qualities among generated proposals.

In summary, we contribute Virtual Dynamic Storyboard (\model):
(1) It takes story and camera scripts as input and translates them into dynamic shot sequences for pre-production that follows cinematic filming  rules.
(2) The design of our cinematic filming module with the associated camera action subspace not only brings interpretable control to users but also effectively reduces the action space which leads to more plausible results.
(3) The shot ranking discriminator mitigates the gap that there are no ground truth and universal criteria by learning from professional manual-created clips with carefully designed class-aware contrastive objectives. 
(4) Extensive user studies and qualitative evaluations demonstrate the effectiveness of \model. 
Codes will be released upon publication.

\vspace{1pt}
\noindent\textbf{Assumption.} 
Though our tool can generate a sequence of shots, its technique novelty mainly focuses on how to compose a single shot.
The users have the freedom to decide how different shots are connected to bring out a sequence and their transition effects. 
Hence, we assume that each shot duration is equal to its corresponding atomic character action and they are connected directly in our paper to form a sequence, 
noting that different shots can be ``cut'' and connected in various ways using our shots as raw data~\cite{pardo2021learning, pardo2021moviecuts}.
This corresponds to the industry process that applies multiple cameras to record one action and take them as raw inputs to the post-editing process with ``cut''.

%% file: articles/related.tex
\section{Related Work}
\label{sec:related}

\noindent\textbf{Storyboard and dynamic content creation.}
Storyboards are investigated in keyframe or text summarization from videos~\cite{mohanta2013novel, bhaumik2015real, ronfard2022prose}, textual script writing~\cite{chandu2019storyboarding,mirowski2022co} and video creation assistance~\cite{goldman2006schematic,ye2008towards,pizzi2010automatic}.
Among the most relevant ones,
Ye~\etal~\cite{ye2008towards} focus on the language mapping between the action descriptions and avatar animation instead of the visual content and PIzzi~\etal~\cite{pizzi2010automatic} only generate sketch-style static images.
Besides, intelligent creation tools are 
in great demand as they can help users efficiently create customized dynamic content, \eg, video, animation~\cite{louarn2018automated,louarn2020interactive}.
Some researchers focus on several key steps, such as frame composition~\cite{zhong2021aesthetic}, shot selection~\cite{liao2020occlusion,jiang2021jointly}, shot cut suggestion~\cite{pardo2021learning}.
Others tackle high-level automatic procedures with simple user interactions
and take multiple videos captured by different cameras to produce a coherent video in different application scenarios~\cite{arev2014automatic,leake2017computational,truong2019tool} 
using different data sources~\cite{wang2019write, chi2021automatic, moorthy2020gazed, rao2022shoot360, rao2022temporal}.
Our system also belongs to high-level automatic creation that takes 
story/camera scripts as input.

\vspace{1pt}
\noindent\textbf{Camera control for cinematography.}
Camera motion plays an important role in delivering content from a given environment~\cite{he1996virtual,wu2018thinking}.
Early works start from handcrafting quality criteria such as visibility and smoothness~\cite{huang2016trip, galvane2015camera} for route planning~\cite{oskam2009visibility}.
However, they are limited in the scope of possible actions and lack generalizability in broader scenarios.
Later works tend to directly imitate exemplar videos
to generate similar camera trajectories with SfM~\cite{sanokho2014camera}.
Jiang~\etal~\cite{jiang2020example} use deep neural networks to extract camera behaviors from real movie clips based on toric space~\cite{lino2015intuitive}, 
which is helpful in imitating camera motion~\cite{yoo2021virtual} and keyframing~\cite{jiang2021example}.
However, the inaccuracy of SfM and toric space estimation may severely affect their performance.
Another research direction targets drone photography~\cite{gebhardt2021optimization,galvane2018directing}, which explores imitation learning to learn from experts. 
Reinforcement learning is used to maximize an aesthetic-based reward~\cite{gschwindt2019can, huang2019cvpr} with extension to style control~\cite{huang2021one}.
However, the shot styles they studied are quite limited compared to the broad categories of plausible shots in real film production, and it is generally not easy to acquire ground truth trajectory data for training.

\vspace{1pt}
\noindent\textbf{Simulation and virtual environments.}
Simulation engines can facilitate the training of machine learning models in autonomous driving, drones, robots and so on~\cite{richter2016playing,brodeur2018home,gao2019vrkitchen}, with the potential to augment an infinite number of data samples.
While many virtual platforms for real-world scenes can only complete some single agent tasks~\cite{starke2019neural,zhang2018mode}, 
a line of research focus on social AI to construct multi-people tasks recently~\cite{savva2019habitat,shridhar2020alfred} and develop virtual scenes are based on Unity~\cite{unity} or Unreal~\cite{unreal}.
With developed APIs, 
our \modelss supports script-level auto control over Unity~\cite{unity} and Omniverse~\cite{omniverse} engines.
It is able to simulate over $100$ kinds of camera shot styles and character actions to fulfill a storyboard, and render it out.

%% file: articles/method.tex
\section{Virtual Dynamic Storyboard}
\label{sec:methodology}

\begin{figure*}[!t]
	\vspace{-2pt}
	\includegraphics[width=\linewidth]{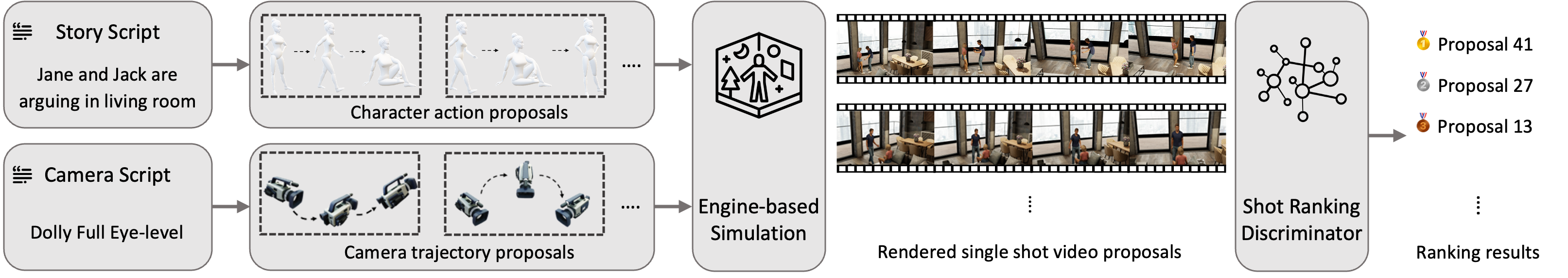}
	\vspace{-20pt}
	\caption{
		The pipeline of Virtual Dynamic Storyboard to process each shot.
		It first generates multiple proposals that match the requirements of the story and camera script and renders videos out with an engine-based simulation module.
		The shot ranking discriminator then scores the generated proposals according to their quality. 
	}
	\label{fig:pipeline}
	\vspace{-10pt}
\end{figure*}

The framework of Virtual Dynamic Storyboard (VDS) is shown in Fig.~\ref{fig:pipeline}. 
It first proposes several candidates of executable parameters for each shot's story script and camera script as inputs (Sec.~\ref{sec:script} and Sec.~\ref{sec:film}), following the story and cinematic rules.
The proposal storyboard videos are then rendered by an engine-based simulator that satisfy the corresponding parameters.
Through class-aware contrastive learning, a shot ranking discriminator helps to score the generated proposals and assists users in selecting their favorites (Sec.~\ref{sec:rank}).
The practical UI is presented in Sec.~\ref{sec:user_interface}.

\subsection{Story Script Proposal}
\label{sec:script}
One of the raw inputs to VDS is a sentence of story scripts in the format of $(\langle$char$\rangle$ $\langle$do$\rangle$ $\langle$sth/swh$\rangle)$.
Although graphic engines have strong ability in scene establishment and character animation with manual force (mouse clicking/dragging and keyboard typing),
these functions cannot be directly called with the aforementioned story scripts. 
To fulfill this goal, we develop an \emph{engine-based simulation module} with a series of automatic APIs including: 
scene selection, character placement and animation, camera control.

After the selection of characters and scenes from available assets,
each story script allows 
a chosen virtual character to interact with a predefined scene 
and executes its animation for each timestamp in the simulator.
Specifically,
1) \emph{$\langle$char$\rangle$ and $\langle$sth/swh$\rangle$:}
$\langle$char$\rangle$ links the selected character asset.
And scene is represented as a hierarchical tree, \eg, a house is composed of several rooms, 
which in turn are composed of several objects. 
From this scene graph, each object and place have their position, 
which can be associated with the $\langle$sth/swh$\rangle$ in the story scripts.
2) \emph{$\langle$do$\rangle$:}
For each atomic action $\langle$do$\rangle$, 
we retrieve corresponding pre-recorded $N$ proposal animation clips $[\va_1, \va_2, \dots, \va_N]$ from a predefined animation clip pool and associate them with the character and object/place. 
Each clip $\va_n$ can be executed by the simulator that output a video reflecting the semantics of atomic motion.
If the atomic action involves a distance-motion, \eg, walk/run, 
$M$ proposal paths $[\vp_{n,1}, \vp_{n,2}, \dots, \vp_{n,M}]$ are generated between the character and the object for $\va_n$. 
Each paths are selected with the scene graph to avoid objects getting in the way. 
The motion trajectory of the character can be accurately represented by
$\vp_{n,m} = (x_{p_{n,m}}^t, y_{p_{n,m}}^t, z_{p_{n,m}}^t), t \in [0, T)$ in the world coordinate,
and $T$ is equal to the length of $\va_n$.

The story parameters $\mathbf{s} \in \cS$ of each executable storyboard animations for one story script can be represented as: 
\begin{equation}
\mathbf{s} = (\va_n, \vp_{n,m}) , \quad \mathbf{s} \in \cS.
\end{equation}

%

\subsection{Camera Script Proposal}
\label{sec:film}

\smallskip
\noindent\textbf{Preliminaries for camera control.}
Following our assumption, each story script $(\langle$char$\rangle$ $\langle$do$\rangle$ $\langle$sth/swh$\rangle)$ will produce $\cS$ executable animations, 
and the filmed shot duration is the same as the character's animation length, which is specified by its corresponding action clip $\va$.
Thus, in order to take a shot for each atomic action,
it is necessary to obtain a list of 7DoF camera parameters that represent the camera trajectory:
\begin{equation}
	\vc = \{(x^t, y^t, z^t), (\alpha^t, \beta^t, \gamma^t), f^t\}, \quad t \in [0, T),  \quad \vc \in \cC,
\end{equation}
where $(x^t,y^t,z^t)$ stands for the position in world coordinates. 
The roll $\alpha^t$, pitch $\beta^t$, and yaw $\gamma^t$ describe the camera rotations along the $Z_C, X_C, Y_C$ axes in the camera local coordinate, where the upward direction 
is specified by a ``Look-At'' constraint.\footnote{The commonly adopted \href{https://docs.unity3d.com/ScriptReference/Transform.LookAt.html}{``Look-At''} constraint for camera is realized by a TR matrix that positions/rotates the camera to look at a target point in space from its positioned point, according to the orientation set up by the up-vector and the looking direction.
	It positions the camera to be horizontal in most cases except for some advanced effects, such as ``roll'' shots.
	Details are specified in the supplementary. 
}
The focal length $f$ is set in the commonly used range $[30, 80]$mm of films.


\begin{figure}[t]
	\centering
	\includegraphics[width=0.9\linewidth]{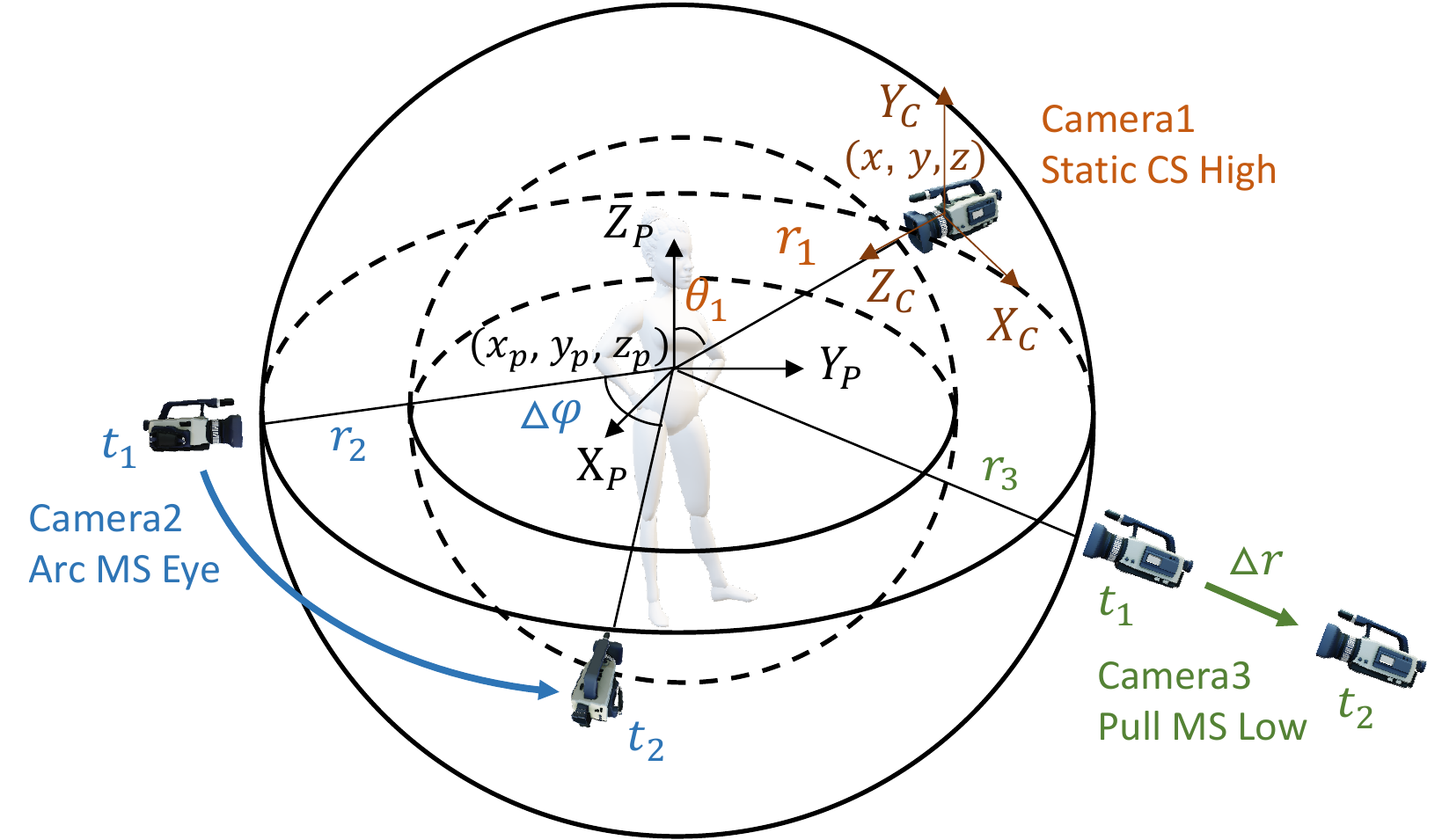}
	\vspace{-9pt}
	\caption{Illustration of camera parameters and shot scale/angle/movement, where three kinds of shots are presented.
		$(x,y,z)$ and $(x_p,y_p,z_p)$ are the camera and character position in the world coordinate. 
		$X_C, Y_C, Z_C$ and $X_P, Y_P, Z_P$ represent the camera and character local coordinate.
	}
	\label{fig:cameras}
	\vspace{-8pt}
\end{figure}

\smallskip
\noindent\textbf{Cinematic camera control.}
To associate the produced shots with the classical cinematic style filming language, and provide easy control to users,
we introduce three dimensions for shot type control and allow the input in the format of $(\langle$movement$\rangle$ $\langle$scale$\rangle$ $ \langle$angle$\rangle)$,
which are widely used in the filming industry to increase production efficiency~\cite{rao2020unified}.
The definition of these three factors and their complete list of subcategories can be found in \cite{giannetti1999understanding}.
Instead of freely searching from the general camera action space,
we identify a set of camera \emph{subspaces} that defines meaningful camera trajectories in terms of shot scale, angle, and movements.

In the following, we adopt a \emph{human-centric} explanation and introduce a direction vector represented in the spherical coordinate system $(r^t,\theta^t,\varphi^t)$ 
in terms of the radial distance $r^t$, polar angle $\theta^t$, and azimuthal angle $\varphi^t$ to facilitate control. 
This representation explicitly derives the relative positions between the main character and the cameras, and bears the merit of having an one-to-one mapping from $(x^t,y^t,z^t)$ to $(r^t,\theta^t,\varphi^t)$ via:
\begin{equation}
	\small{
		(x^t , y^t , z^t )  = (x_{p}^t , y_{p}^t , z_{p}^t ) +  r^t (\cos \varphi^t  \,\sin \theta^t , \sin \varphi^t  \,\sin \theta^t ,\cos \theta^t  ).
	}
\end{equation}
Based on the specification of the shot scale and angle, 
we can determine the camera parameters for the keyframes in a shot, 
while the type of camera movement finally determines the camera parameter for each interpolated timestamp $t$ in-between $[0, T)$.
Fig.~\ref{fig:cameras} shows an illustration of the camera parameters and a selection of shot types.
Some basic categories are shown below, which can be naturally extended to more with details explained in the supplementary.

\subsubsection{Shot Angle}
This is largely determined by the relative position between camera and target, with a special focus put on their altitude difference. As shown in Fig.~\ref{fig:cameras}, different angles are reflected by the vector $\theta$,~\eg, $\theta = \pi/2$ represents \textbf{eye-level} shot, $\theta = 2\pi/5$ produces \textbf{high-angle} shot, and $\theta = 4\pi/5$ serves \textbf{low-angle} shot.

\subsubsection{Shot Scale}
It is determined by the size of the target object within the frame, 
which is implemented with a virtual sphere centered on the target character with  the distance radius $r^t$
and the focal length $f^t$.
For example, for $f^t=50$~mm and a character with height $h$,
$r^t \in \{0.2h,0.5h, h\}$ is pre-set
for \textbf{close-ups}, \textbf{medium}, \textbf{full} shots respectively.
$f^t$ and $r^t$ can be adjusted accordingly to meet individual needs.

\subsubsection{Shot Movement} 
In the human-centric scenario, the camera movements are correlated with the motion trajectory of the character.
Based on the above controls for shot scale and angle,  we elaborate the control of camera movements
that depicts the camera parameters for $\forall t \in [0,T)$. 
Nine basic movement types are briefly explained below.

\noindent\textbf{Static} shot
keeps the 7DoF as constants throughout the time duration with the reference target coordinate as 
the character's start  $(x_p^0, y_p^0, z_p^0)$ or end $(x_p^{T-1}, y_p^{T-1}, z_p^{T-1})$ position.

\vspace{1pt}
\noindent\textbf{Follow} shot aims to follow the character's motion trajectory within the time duration $T$, and keeps looking-at the character. 
With an easing function $w_{\lambda}(t)$ parameterized by $\lambda$ to control the movement rhythm, the camera position at time $t \in [0,T)$ is determined by,
\vspace{-2pt}
\begin{equation}
	\small{
		\begin{aligned}
			(x^{t}, y^t, z^t)&  = (x_p^{w_\lambda(\frac{t}{T-1})}, y_p^{w_\lambda(\frac{t}{T-1})},z_p^{w_\lambda(\frac{t}{T-1})}) \\
			&+ (r\cos \varphi \,\sin \theta, r\sin \varphi \,\sin \theta,r\cos \theta ),
		\end{aligned}
	}
\end{equation}
\vspace{-5pt}
\begin{equation}
	\small{
		\begin{aligned}
			\quad w_{\lambda}(t) &=\left\{
			\begin{aligned}
				&\frac{\lambda^t-1}{\lambda-1}, &\lambda \in (0,1) \cup (1, \infty), \\
				&t,  &\lambda = 1.
			\end{aligned}\right.
		\end{aligned}
		\label{eq:easing}
	}
\end{equation}
In general, a large $\lambda$ makes the shot ``slow first, fast later'' and vice versa. 
The camera rotation parameters are determined by the Look-At constraint.

\noindent\textbf{Push/pull} shot compresses/enlarges the shooting space to focus on a single object or show the surroundings. 
It adjusts the distance $r$ between the camera and the subject with a zoom ratio parameter $\mu$ and easing function $w_{\lambda}(t)$ in Eq.~\eqref{eq:easing},
\begin{equation}
	\small{
		r^t = ((\mu -1) w_{\lambda}(t)+1) r^0.
	}
\end{equation}
\noindent\textbf{Zoom} shot are similarly achieved by adjusting the camera focal length and its scale and angle are defined based on the first frame.
\textbf{Tilt} shot and \textbf{pan} shot rotate the pitch angle $\beta$ or the yaw angle $\gamma$ respectively to shift audience's focus from one to the other vertically or horizontally.
\textbf{Dolly} 
(horizontal) and \textbf{pedestal} (vertical) shots are implemented by specifying a camera trajectory
based on starting and ending points with the easing function, which is similar to push/pull.
\textbf{Arc} shot is usually applied to a character staying at a specific location, 
where the camera moves around the subject rotating the azimuthal angle $\varphi$ in the direction vector. 

The above definition constrain some variables in $\vc$ depending on the shot angle, scale and movement types.
Then we enumerate the unconstrained variables within their ranges (a subspace of 7DoF) to acquire camera trajectory proposals.

\subsection{Shot Ranking Discriminator}
\label{sec:rank}
For a given story and camera script, the subspaces defined above can produce multiple $\cS \cdot \cC$ plausible shot proposals rendered by our simulator differing in their specific parameters,~\eg, character paths and camera views. 
The next question is how to effectively evaluate these shots and select the optimal shots that look best 
in the combination of content, environment, and proposed camera trajectory.
Since there are no standard metrics for evaluating a good shot, and simple criteria such as smoothness are quite limited in telling the overall shot quality,
we propose a data-driven \emph{shot ranking discriminator} to score the quality of generated shots,
such that users can easily select high-quality shots based on the scores.
To acquire more capacity in discriminating the  spatial-temporal structure among shot proposals,
it is implemented with a TSN structure~\cite{wang2018temporal}  that samples 8 images from a shot of size 224 × 224, and subsequently feeds each image to a ResNet50 and fuses extracted features in the end.

We train the binary classifier using the professional manual-created clips as positive samples and the randomly generated virtual samples as negative ones,
under the assumption that the professional manual-created clips represent higher quality.
The network is trained with the loss,
\vspace{-2pt}
\begin{equation} 
	\small{
		\cL_b = - y \log (p_b) + (1-y) \log(1-p_b).
	}
\end{equation}
During inference, the generated samples can be sorted by their classification scores for being categorized as ``professional''. 
And our goal is to find the one with the highest score that fools the network into treating them as professional samples. 

Nevertheless, 
we found that 
the network could only learn superficial appearance-level criteria to distinguish two classes,~\eg, color/texture
due to the shot type variance.
The features of the generated samples to be ranked also stick to each other in the feature space, 
and it is less informative to distinguish their quality in terms of likeness towards high-quality shots.
To overcome this problem, we design the following two objectives to facilitate the pick-up of high-quality shots. 

\vspace{1pt}
\noindent \textbf{Class-aware contrastive objectives.}
Considering the variance among generated samples caused by their intrinsic shot styles,
we add a \emph{class-aware} loss to encourage it to be more class type specific and be expert in determining the corresponding shot type's quality:
\vspace{-10pt}
\begin{equation} 
	\small{
		\cL_c = - \sum_{c=1}^{M} y \log (p_c),
	}
\end{equation}
where $M$ is the total number of shot types in training data.

To better select the high-quality generated shots,
we need to magnify the feature difference among samples in different quality.
Inspired by recent successes in video contrastive learning~\cite{pan2021videomoco} that could learn high-level features for each shot, such as layout, temporal pace,
which is helpful to determine the generated quality, 
we propose to include a \emph{contrastive} objective,
\vspace{-5pt}
\begin{equation}
	\small{
		\cL_q = -\log\frac{\exp{(z_q z_+ /\tau)}}{\sum_{1}^{K} \exp{( z_q z_k /\tau)}}.
	}
\end{equation}
The loss is computed over one sample $z_q$ and $K$ other samples $\{z_k\}$
in the training set.
A large $K$ means that more shots are taken into account when maximizing the difference among shot clips which leads to better performance. 
$z_{+}$ comes from different frames at other timestamps within the same shot as the target sample $z_q$.

In summary, 
at the training time, 
the shot ranking discriminator has been optimized with the following composite loss, with the aim of figuring out the high-quality ones from multiple shots: $\cL = \cL_b + \cL_c + \cL_q .$
We take $5,000$ professional manual-created clips as positive samples, and $5,000$ randomly generated shots as negative samples. 
For the positive samples, 9 professional designers select the \modelss generated samples in high quality with double checked by each other.
The negative samples come from a random perturbation in the camera action space.
At the inference time, the shot ranking discriminator sorts $\cS \cdot \cC$ proposals of each shot according to their classification scores, 
and the sample with the highest positive score $p_b$ will be chosen as the final output.

\subsection{Practical User Interface}
\label{sec:user_interface}

\begin{figure*}[!t]
	\includegraphics[width=\linewidth]{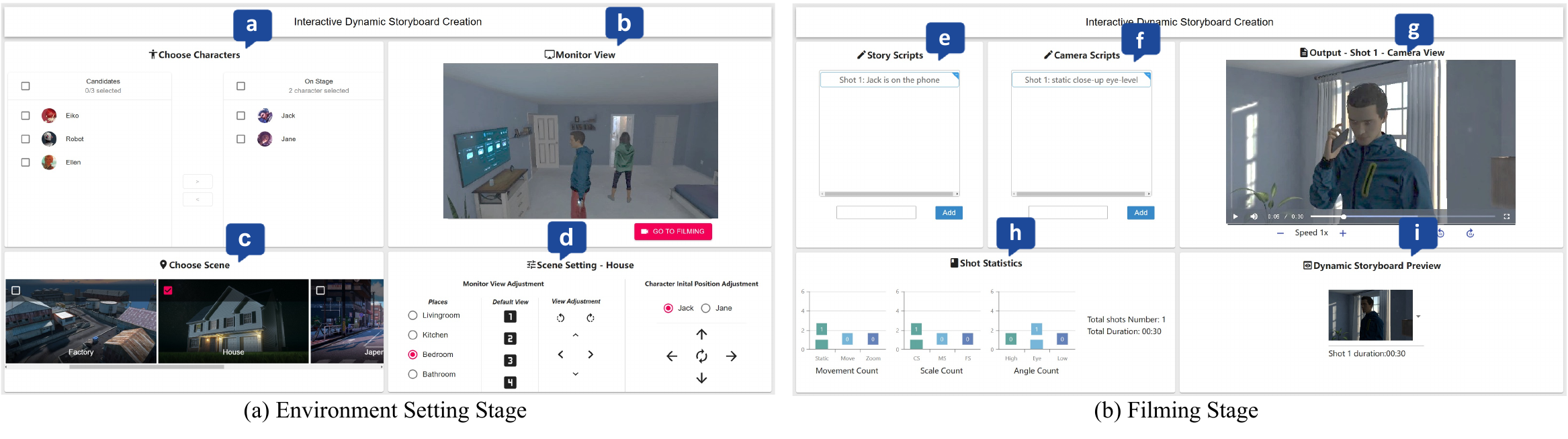}
	\vspace{-24pt}
	\caption{
		The two-stage user interface of \model. 
		(a) Environment Setting Stage: Users select where the story takes place,~\eg, a house, add desired characters to it and adjust their position and orientation. 
		(b) Filming Stage: Users input the story scripts and camera scripts and then acquire the generated dynamic storyboard. 
		Users are free to delete, add, and change scripts at any time to change the generated results in a flexible and convenient way.
	}
	\label{fig:interface}
	\vspace{-10pt}
\end{figure*}

As a pre-production tool designed for the practical video production pipeline, 
Virtual Dynamic Storyboard is used to rehearse the plot and provide a guide in the downstream on-stage videography.
To fulfill this goal, we resort to the advice from professional storyboard staff and 
present a two-stage user interface prototype including Environment Setting Stage and Filming Stage. 
In Fig.~\ref{fig:interface}, we show a brief introduction of it with 
alphabets highlights of each panel. 
The key design idea is to divide the storyboard creation into static and dynamic stages, that allow users  to apply clearer controls on the static scenes, dynamic characters and cameras.

Following this idea, 
the first step is to prepare scene and character assets for the story and camera scripts in a static Environment Setting Stage.
\emph{Choose Scene window (a)} and \emph{Choose Characters window (b)} allow user to select the characters and their initial locations in story. 
Users can freely use \emph{Scene Setting window (d)} to control the viewing angle of the monitor camera to facilitate the operation of the scene and view the results through \emph{Monitor View window (c)}.

With the chosen assets, the system is ready to produce dynamic storyboard in the Filming Setting Stage.
\emph{Input windows (e, f)} provide users with text input to the system. 
Users can type story scripts and camera script to check the results in the \emph{Output window (g)} of the corresponding story.
To monitor the generated results timely, the keyframe of each shot can be found in the \emph{Preview window (i)}.
Users are allowed to click the drag box and select more cases according to the ranking score.
Additionally, in the \emph{Statistics window (h)}, 
we visualize some basic real-time statistics of the generated storyboards, \eg, the shot style counting, total shot number, etc.

%% file: articles/experiment.tex
\section{Experiments}
\label{sec:experiment}

\subsection{Implementation Details}

The assets used in \modelss contain over $20$ different scenes, $100$ actions and $20$ characters.
The character height $h$ in the simulation engine is set as $1.6m$.
Each story and camera script pair obtains 40$\sim$200 proposals. 
For a fixed shot scale, we generate proposals with $8$ different azimuthal angles to represent different shooting directions.
For the easing function, we enumerate $\lambda \in \{0.1, 1, 10\}$.
In push/zoom-in shots, $\mu$ is set to  $0.5\sim0.8$ 
and for pull/zoom-out shots, $\mu$ is set to  $1.0\sim1.2$ .
In tilt and pan shots, the angle change is constrained within 30$^{\circ}$$\sim$60$^{\circ}$, and we enumerate all the combinations of the camera moving directions (\eg up/down), ending points (on/off the person).
In arc shots, the azimuthal angle changes within 90$^{\circ}$$\sim$120$^{\circ}$.
We perform the inference of \modelss on a laptop with an Intel i7 CPU
and an NVIDIA 2080Ti GPU to output 720P videos and the processing time is shown in Tab.~\ref{tab:process}.
The training process of shot ranking discriminator takes batch size 128, $1\mathrm{e}{-3}$ as the learning rate with $60$ training epochs.
To compute $\cL_q$, we used the momentum updated dictionary~\cite{he2020momentum} and set the dictionary size as $6553$ corresponding to $K$.
And $\cL_c$ is the sum of shot movement, scale, angle classification losses. 
More details can refer to the supplementary.

\subsection{User Evaluation on Storyboard Designer}
\label{subsec:design}

\noindent\textbf{Setting.}
To evaluate the performance of \modelss in practical usage, 
we invite 20 amateur designers to compare the usage of hand-paint and our \modelss. 
8 of them are \textbf{skilled users} who learn painting over two years while the rest 12 of them are \textbf{beginners} without painting skills.
Before the experiments, 
they take a 30-minute training session to familiarize with the tool
and read reference material showing some examples of the types of activities and characters.
After the training, 
each of them first uses traditional hand-paint to create storyboards for 2 different practical stories. 
Each story depicts a scene containing $\sim$10 shots.
Then they utilize \modelss to generate top 5 results for each shot and pick up one to create the dynamic storyboards on the same stories.
In the process,
we count their creation time to study the time efficiency,
and ask them to pair-wised self-evaluate which approach can generate better results that reach their expectation.
In the end,
we ask them to vote which method holds higher flexibility in creation in Tab.~\ref{tab:designer}.

\begin{table}[!t]
	\begin{center}
		\caption{Processing speed statistics of each step.}
		\label{tab:process}
		\vspace{-10pt}
		\resizebox{0.75\linewidth}{!}{
			\begin{tabular}{c|cccc}
				\toprule
				step   & proposal & render &	shot ranking 
				\\ 
				\midrule
				speed	 &    24.8 item/s  & 13.0 frame/s  &  10.3 shot/s    \\
				\bottomrule    
			\end{tabular}
		}
		\vspace{-5pt}
	\end{center}
\end{table}

\begin{figure*}[!t]
	\vspace{10pt}
	\includegraphics[width=\linewidth]{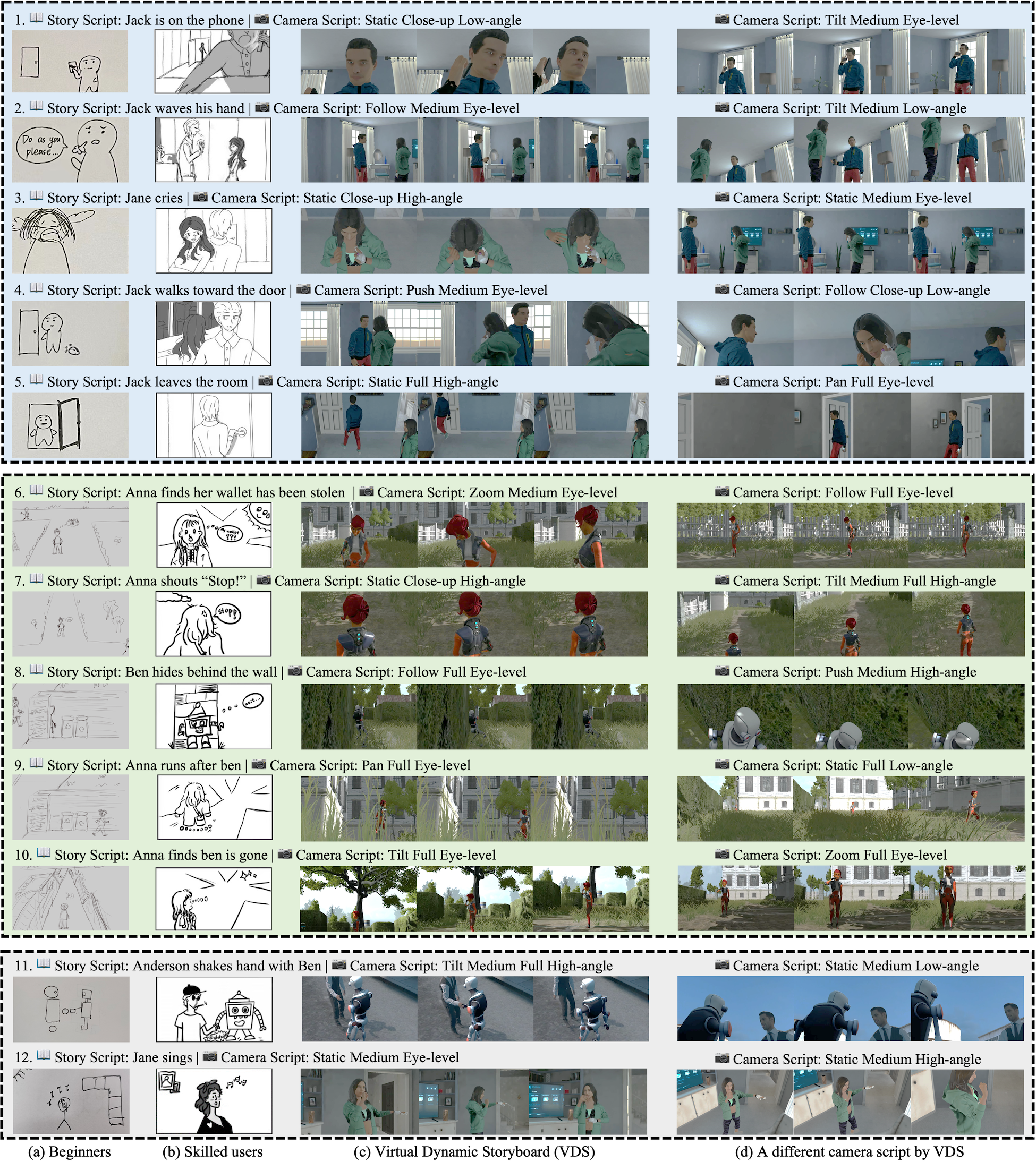}
	\vspace{-19pt}
	\caption{
		Qualitative comparison among (a) beginners hand-paint, (b) skilled users hand-paint and (c) results generated by \modelss and each shot is represented with three keyframes.
		(d) shows the results sharing the same story script with a different camera script.
		The 1-5 and  6-10 rows are five consecutive scripts corresponding to two scenes.
	}
	\label{fig:compar}
\end{figure*}

\vspace{1pt}
\noindent\textbf{Results.}
Fig.~\ref{fig:compar} compares the storyboard generated by hand-paint and \modelss with the same story and camera script in an indoor and an outdoor scene respectively.
Beginners without painting skills have difficulties in expressing the story with the camera script, \eg, the character just stands in front of door in the 5-th row of Fig.~\ref{fig:compar}~(a). 
The skilled users tend to use simple shapes and omit details in the backgrounds to design, \eg, the road is depicted by some simple lines in the 9-th row of Fig.~\ref{fig:compar}~(b). 
For actions, hand-paint also tends to use some symbols, \eg, uses musical notes to indicate that the character is singing in the 12-th row of Fig.~\ref{fig:compar}~(b). 
It is observed that hand-paint requires painting skills for designers and more time cost to create a better storyboard.
Thanks to \model's support for rich action and scenes, 
amateurs do not need painting skills to create a dynamic storyboard. 
And it also provides easier modification and more options for users as shown in Fig.~\ref{fig:compar}~(d).

\begin{table}
	\begin{center}
		\caption{Pair-wise comparison of hand-paint and \modelss
			on time cost per shot in minutes and 
			binary voting.
		}
		\label{tab:designer}
		\vspace{-12pt}
		\resizebox{\linewidth}{!}{
			\begin{tabular}{ll|cccc}
				\toprule
				\makecell[c]{Designer\\Type}&\makecell[c]{Approach}&  \makecell[c]{Time Cost\\ per shot} & \makecell[c]{Reach \\ Expectation} &  \makecell[c]{Camera \\ Flexibility}   & \makecell[c]{Story \\ Flexibility}  \\ 
				\midrule
				\multirow{2}[1]{*}{Beginner} &  Hand-paint & 6.69 min    & 16.67\%   & 8.33\%         & \textbf{75.00\%}             \\
				& \model  & \textbf{3.42} min   & \textbf{83.33\%}      &  \textbf{91.67\%}    &25.00\%        \\ 
				\midrule \midrule
				\multirow{2}[1]{*}{Skilled user} &  Hand-paint & 12.73 min      & 25.00\%  & 12.50\%          & \textbf{62.50\%}             \\
				& \model  & \textbf{3.13} min       & \textbf{75.00\%}      &  \textbf{87.50\%}         &37.50\%    \\ 
				\bottomrule
			\end{tabular}
		}
	\end{center}
	\vspace{-5pt}
\end{table}

Tab.~\ref{tab:designer} shows that 
\modelss provides a lower creation time cost of 3 minutes per shot to reach their expectation.
And they vote for its high flexibility on camera settings.
Compared to the hand-paint that can freely draw anything,
\modelss sacrifices some of its story flexibility.
To have a deeper understanding of their choices,
we conduct a semi-structured interview with them.
The designers enjoyed the efficiency, performance and rich camera choices brought by \modelss.
Though the new tool sacrifices a bit of flexibility on story design,
11 out 20 of them mentioned that this is relevant to their painting skills under different scenarios.
``When I deal with common cases,~\eg, a couple talking in a bedroom, 
hand-paint allows me to design freely.
But if I want to design some things in a different style,~\eg, the magic realism like \emph{The Lord of the Rings} or Postmodernism like Westworld,
the provided assets ease our design process.''
Based on the above observation,
we believe that as the development of more assets, \modelss will provide more flexibility when we put more available assets in it.
And it also opens up opportunities for those people without painting skills to design their dynamic storyboards.

\vspace{-1pt}
\subsection{User Evaluation on Storyboard Reader}

\noindent\textbf{Setting.}
The most important functionality of storyboard is to convey the ideas of creators and guide the videography process.
To validate this, 
we invite 22 amateurs as storyboard readers, which be splitted into 3 groups (8,10,4) according to the year they use storyboard in video production.

Facing the storyboards created by the above designers in Sec.~\ref{subsec:design},
we ask the readers whether they can understand the content in the storyboard and get the information for filming.
To further understand which aspects play important roles, 
we conduct pair-wise rating on the elements of story and camera, where we set the hand-paint as $4$ and asks them to rate corresponding \modelss results in seven-point Likert scale.

\begin{table}[!t]
	\begin{center}
		\caption{Pair-wise comparing to hand-paint,
			the satisfying rating of \modelss on different aspects in percentage and seven-point Likert scale (lowest-highest: 1-7).
		}
		\label{tab:user}
		\vspace{-10pt}
		\resizebox{0.94\linewidth}{!}{
			\begin{tabular}{l|c|c|c}
				\toprule
				Year use storyboard/\# users &  0$\sim$0.5 / 8                      &  0.5$\sim$1.5  /10          &  1.5$\sim$2.5  / 4                               \\\midrule
				Content Delivery                             & {72.50\%}                         & {62.00\%}                                             & {95.00\%}                        \\
				Instruction Ability    & {77.50\%}      & {80.00\%}        & {100.00\%}                  \\ 
				\midrule	\midrule
				Story Character              & 5.45 $\pm$ 0.53                               &   5.36   $\pm$ 0.37                         &  6.25     $\pm$ 0.50                         \\
				Story Action                     & 5.25          $\pm$ 0.68                                & 5.50      $\pm$ 0.95                                 & 6.70      $\pm$ 0.47                       \\
				Story Scene             & 5.30         $\pm$ 0.38                              & 5.12    $\pm$ 0.42               & 6.20 $\pm$ 0.80                           \\
				\midrule
				Camera Scale                       & 4.98     $\pm$ 0.32                             & 4.36 $\pm$ 0.70                                    & 6.05     $\pm$ 0.81                       \\
				Camera Angle                             & 4.58     $\pm$ 0.59                                    & 4.02      $\pm$ 0.69                           & 6.80     $\pm$ 0.71                             \\               
				Camera  Movement              & 5.67  $\pm$ 0.39                                 & 5.18    $\pm$ 0.49                                  &  6.85     $\pm$ 0.51                        \\
				\bottomrule
			\end{tabular}
		}
	\end{center}
	\vspace{-10pt}
\end{table}

\vspace{1pt}
\noindent\textbf{Results.}
Tab.~\ref{tab:user} shows that 
the results produced by \modelss reach significantly better performance in content delivery and instruction ability,
with above 70\% for readers without much experience, and above 95\% for people with $\sim$2-year experience.
Readers with more experiences appreciate its better incorporation and usefulness in the practical video production pipeline.
As we dive deeper into the affecting components behind it,
it is found that people prefer the story setting of \modelss,~\ie, character, action and scene. 
As mentioned above, hand-painted storyboards tend to be very simple and simplify the characters, actions, and scenes to a great extent. 
Only those designers who are skilled in painting can accurately convey their ideas with the hand-paint. 
In contrast, the dynamic storyboards generated with \modelss make it easy for users to understand the information in the story. 
For camera movement, users prefer \modelss too.
Still images are unfriendly for users to restore the camera motion. 
Instead, \modelss produces more intuitive results, and users can directly see how the frames proceed.

To have a deeper understanding of \modelss and its comparison with conventional hand-paint,
we conduct a semi-structured interview with participants and collect the reasons why they prefer or dislike \modelss or hand-paint.
They feel enthusiastic about the ability offered by \modelss to speed up their creation procedure, 
and consider this tool a good substitute for the storyboard, which is more descriptive and vivid.
Among the 22 storyboard users,
16 of them prefer \modelss 
due to its dynamics which let the shot to be much easier to understand. 
19 of the participants appreciate its standardized process that is similar to the real-world videography.
Especially for the skilled users, they give the highest scores and praise the potential towards a more standardized pipeline for the whole video production.

We also receive feedback on the drawbacks and 
suggestions from them such as ``..., looking forward to adding more characters into the tool, ...'', ``.., it would be better if it could support more characters facial expressions''.
These comments reveal insightful and exciting directions for future improvements to our system.

\subsection{Ablation Study on Shot Ranking}

\begin{figure}
	\centering
	\includegraphics[width=\linewidth]{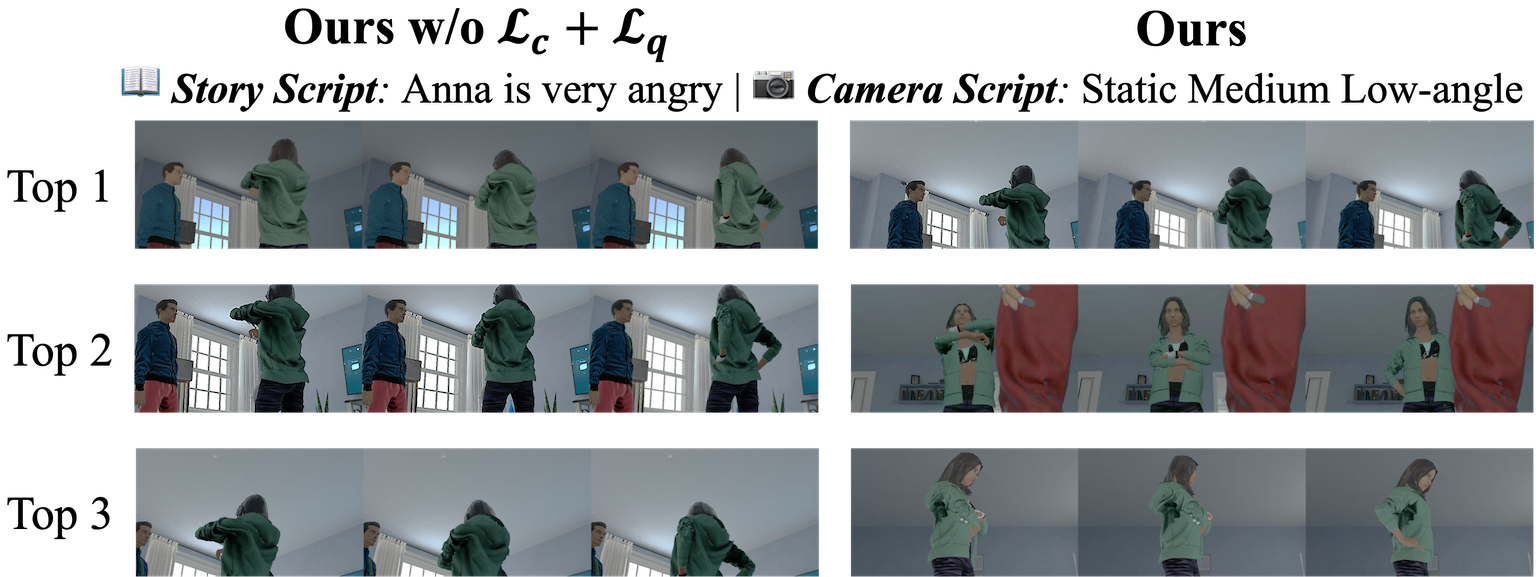}
	\vspace{-18pt}
	\caption{Visualization of top 3 samples of different losses.
		The class-aware loss $\cL_c$ brings out more accurate shot types matched with the camera scripts.
		The constrastive loss $\cL_q$ separates the feature space of shots to be more distinguishable and provide enough diversity in the top results for users to pick up.
		Each shot is represented with three keyframes.}
	\vspace{-10pt}
	\label{fig:multiple}
\end{figure}

To show the rationality of our shot ranking discriminator, 
we additionally collect 200 shots picked by professional designers from \modelss proposals and score them 
with 20,000 generated proposals by the ranking network together.
All designers' shots are ranked in the top 10\% which shows that our shot ranking discriminator is able to pick up high-quality shots.
We also demonstrate the effectiveness of the loss function for training the shot ranking discriminator by visualizing the top 3 shot candidates of fixed scripts (Fig.~\ref{fig:multiple}).
The top 3 results from Ours w/o $\cL_c + \cL_q$ look similar and their shot types are not accurate. 
\eg, the shot scale of Anna does not perfectly match the medium type.
With the help of class-aware contrastive objectives,
the full model is able to display more accurate shot types matched with the input scripts, 
and the top 3 results provide enough diversity for users to pick up.
This is matched with our expectation, since the ablated model without the presence of $\cL_c + \cL_q$ has difficulty in learning a class-aware distinguishable feature space for shot proposals.

%% file: articles/conclusion.tex
\section{Discussion and Conclusion}
\label{sec:conclusion}
Virtual Dynamic Storyboard 
(\modelss) is a semi-auto storyboard creation tool performing on virtual environments.
It takes user-specified story script and camera script as inputs
and proposes multiple plausible character and camera trajectories that can be rendered to be dynamic shots.
A shot ranking discriminator then sorts these generated shots with class-aware contrastive objectives to output the top results.
Experiments show that our tool can effectively compose dynamic storyboards and assist amateurs in their creation.
It also bears the following limitations.

\vspace{1pt}
\noindent\textbf{Quality of assets.}
The quality of \modelss's results heavily depends on the quality of assets,~\ie, character model, action animation and virtual environment.
For example, \modelss can not generate rich expressions for a character if this character asset itself lacks detailed modeling on the face.
As our design is easy to extend with different assets, 
we will continue to improve the quality of \modelss with more advanced 3D assets including characters, associated actions, prefabs and scenes. 
There are also opportunities to incorporate character/motion synthesis~\cite{wang2021scene,hong2022avatarclip}.

\vspace{1pt}
\noindent\textbf{Trade-off of flexibility and efficiency on control.}
\modelss provides a relatively flexible and fast way to generate dynamic storyboards for pre-production, 
but it does not yet allow users to control everything as they do with a pen.
Our current version of \modelss addresses the most important steps in setting a shot,~\ie, the story plot and the shot cinematic styles.  
For different user groups, the interaction portal for control needs to be adjusted to meet their requirements and preferences.
It is promising to provide more detailed control on character and camera for professionals,~\eg, details of characters, lens aperture, though it would trigger more creation time cost. 
And for novice users, it would be more friendly to provide more fuzzy input control. 

%% file: main.bbl

\begin{thebibliography}{61}


\ifx \showCODEN    \undefined \def \showCODEN     #1{\unskip}     \fi
\ifx \showDOI      \undefined \def \showDOI       #1{#1}\fi
\ifx \showISBNx    \undefined \def \showISBNx     #1{\unskip}     \fi
\ifx \showISBNxiii \undefined \def \showISBNxiii  #1{\unskip}     \fi
\ifx \showISSN     \undefined \def \showISSN      #1{\unskip}     \fi
\ifx \showLCCN     \undefined \def \showLCCN      #1{\unskip}     \fi
\ifx \shownote     \undefined \def \shownote      #1{#1}          \fi
\ifx \showarticletitle \undefined \def \showarticletitle #1{#1}   \fi
\ifx \showURL      \undefined \def \showURL       {\relax}        \fi
\providecommand\bibfield[2]{#2}
\providecommand\bibinfo[2]{#2}
\providecommand\natexlab[1]{#1}
\providecommand\showeprint[2][]{arXiv:#2}

\bibitem[Arev et~al\mbox{.}(2014)]%
        {arev2014automatic}
\bibfield{author}{\bibinfo{person}{Ido Arev}, \bibinfo{person}{Hyun~Soo Park},
  \bibinfo{person}{Yaser Sheikh}, \bibinfo{person}{Jessica Hodgins}, {and}
  \bibinfo{person}{Ariel Shamir}.} \bibinfo{year}{2014}\natexlab{}.
\newblock \showarticletitle{Automatic editing of footage from multiple social
  cameras}.
\newblock \bibinfo{journal}{\emph{ACM Transactions on Graphics (TOG)}}
  \bibinfo{volume}{33}, \bibinfo{number}{4} (\bibinfo{year}{2014}),
  \bibinfo{pages}{1--11}.
\newblock


\bibitem[Bhaumik et~al\mbox{.}(2015)]%
        {bhaumik2015real}
\bibfield{author}{\bibinfo{person}{Hrishikesh Bhaumik},
  \bibinfo{person}{Siddhartha Bhattacharyya}, \bibinfo{person}{Mausumi~Das
  Nath}, {and} \bibinfo{person}{Susanta Chakraborty}.}
  \bibinfo{year}{2015}\natexlab{}.
\newblock \showarticletitle{Real-time storyboard generation in videos using a
  probability distribution based threshold}. In \bibinfo{booktitle}{\emph{2015
  Fifth International Conference on Communication Systems and Network
  Technologies}}. IEEE, \bibinfo{pages}{425--431}.
\newblock


\bibitem[Brodeur et~al\mbox{.}(2018)]%
        {brodeur2018home}
\bibfield{author}{\bibinfo{person}{Simon Brodeur}, \bibinfo{person}{Ethan
  Perez}, \bibinfo{person}{Ankesh Anand}, \bibinfo{person}{Florian Golemo},
  \bibinfo{person}{Luca Celotti}, \bibinfo{person}{Florian Strub},
  \bibinfo{person}{Jean Rouat}, \bibinfo{person}{Hugo Larochelle}, {and}
  \bibinfo{person}{Aaron Courville}.} \bibinfo{year}{2018}\natexlab{}.
\newblock \showarticletitle{HoME: a Household Multimodal Environment}. In
  \bibinfo{booktitle}{\emph{International Conference on Learning
  Representations Workshop}}.
\newblock


\bibitem[Chandu et~al\mbox{.}(2019)]%
        {chandu2019storyboarding}
\bibfield{author}{\bibinfo{person}{Khyathi Chandu}, \bibinfo{person}{Eric
  Nyberg}, {and} \bibinfo{person}{Alan~W Black}.}
  \bibinfo{year}{2019}\natexlab{}.
\newblock \showarticletitle{Storyboarding of recipes: grounded contextual
  generation}. In \bibinfo{booktitle}{\emph{Proceedings of the 57th Annual
  Meeting of the Association for Computational Linguistics}}.
  \bibinfo{pages}{6040--6046}.
\newblock


\bibitem[Chi et~al\mbox{.}(2021)]%
        {chi2021automatic}
\bibfield{author}{\bibinfo{person}{Peggy Chi}, \bibinfo{person}{Nathan Frey},
  \bibinfo{person}{Katrina Panovich}, {and} \bibinfo{person}{Irfan Essa}.}
  \bibinfo{year}{2021}\natexlab{}.
\newblock \showarticletitle{Automatic Instructional Video Creation from a
  Markdown-Formatted Tutorial}. In \bibinfo{booktitle}{\emph{The 34th Annual
  ACM Symposium on User Interface Software and Technology}}.
  \bibinfo{pages}{677--690}.
\newblock


\bibitem[Fabbri et~al\mbox{.}(2021)]%
        {fabbri2021motsynth}
\bibfield{author}{\bibinfo{person}{Matteo Fabbri}, \bibinfo{person}{Guillem
  Bras{\'o}}, \bibinfo{person}{Gianluca Maugeri}, \bibinfo{person}{Orcun
  Cetintas}, \bibinfo{person}{Riccardo Gasparini},
  \bibinfo{person}{Aljo{\v{s}}a O{\v{s}}ep}, \bibinfo{person}{Simone
  Calderara}, \bibinfo{person}{Laura Leal-Taix{\'e}}, {and}
  \bibinfo{person}{Rita Cucchiara}.} \bibinfo{year}{2021}\natexlab{}.
\newblock \showarticletitle{Motsynth: How can synthetic data help pedestrian
  detection and tracking?}. In \bibinfo{booktitle}{\emph{Proceedings of the
  IEEE/CVF International Conference on Computer Vision}}.
  \bibinfo{pages}{10849--10859}.
\newblock


\bibitem[Galvane et~al\mbox{.}(2015)]%
        {galvane2015camera}
\bibfield{author}{\bibinfo{person}{Quentin Galvane}, \bibinfo{person}{Marc
  Christie}, \bibinfo{person}{Chrsitophe Lino}, {and} \bibinfo{person}{R{\'e}mi
  Ronfard}.} \bibinfo{year}{2015}\natexlab{}.
\newblock \showarticletitle{Camera-on-rails: automated computation of
  constrained camera paths}. In \bibinfo{booktitle}{\emph{Proceedings of the
  8th ACM SIGGRAPH Conference on Motion in Games}}. \bibinfo{pages}{151--157}.
\newblock


\bibitem[Galvane et~al\mbox{.}(2018)]%
        {galvane2018directing}
\bibfield{author}{\bibinfo{person}{Quentin Galvane},
  \bibinfo{person}{Christophe Lino}, \bibinfo{person}{Marc Christie},
  \bibinfo{person}{Julien Fleureau}, \bibinfo{person}{Fabien Servant},
  \bibinfo{person}{Fran{\c{}} ois-louis Tariolle}, {and}
  \bibinfo{person}{Philippe Guillotel}.} \bibinfo{year}{2018}\natexlab{}.
\newblock \showarticletitle{Directing cinematographic drones}.
\newblock \bibinfo{journal}{\emph{ACM Transactions on Graphics (TOG)}}
  \bibinfo{volume}{37}, \bibinfo{number}{3} (\bibinfo{year}{2018}),
  \bibinfo{pages}{1--18}.
\newblock


\bibitem[Gao et~al\mbox{.}(2019)]%
        {gao2019vrkitchen}
\bibfield{author}{\bibinfo{person}{Xiaofeng Gao}, \bibinfo{person}{Ran Gong},
  \bibinfo{person}{Tianmin Shu}, \bibinfo{person}{Xu Xie}, \bibinfo{person}{Shu
  Wang}, {and} \bibinfo{person}{Song-Chun Zhu}.}
  \bibinfo{year}{2019}\natexlab{}.
\newblock \showarticletitle{VRKitchen: An interactive 3D environment for
  learning real life cooking tasks}. In \bibinfo{booktitle}{\emph{International
  Conference on Machine Learning Workshop}}.
\newblock


\bibitem[Gebhardt and Hilliges(2021)]%
        {gebhardt2021optimization}
\bibfield{author}{\bibinfo{person}{Christoph Gebhardt} {and}
  \bibinfo{person}{Otmar Hilliges}.} \bibinfo{year}{2021}\natexlab{}.
\newblock \showarticletitle{Optimization-based User Support for Cinematographic
  Quadrotor Camera Target Framing}. In \bibinfo{booktitle}{\emph{Proceedings of
  the 2021 CHI Conference on Human Factors in Computing Systems}}.
  \bibinfo{pages}{1--13}.
\newblock


\bibitem[Giannetti and Leach(1999)]%
        {giannetti1999understanding}
\bibfield{author}{\bibinfo{person}{Louis~D Giannetti} {and}
  \bibinfo{person}{Jim Leach}.} \bibinfo{year}{1999}\natexlab{}.
\newblock \bibinfo{booktitle}{\emph{Understanding movies}}.
  Vol.~\bibinfo{volume}{1}.
\newblock \bibinfo{publisher}{Prentice Hall Upper Saddle River, New Jersey}.
\newblock


\bibitem[Goldman et~al\mbox{.}(2006)]%
        {goldman2006schematic}
\bibfield{author}{\bibinfo{person}{Dan~B Goldman}, \bibinfo{person}{Brian
  Curless}, \bibinfo{person}{David Salesin}, {and} \bibinfo{person}{Steven~M
  Seitz}.} \bibinfo{year}{2006}\natexlab{}.
\newblock \showarticletitle{Schematic storyboarding for video visualization and
  editing}.
\newblock \bibinfo{journal}{\emph{Acm transactions on graphics (tog)}}
  \bibinfo{volume}{25}, \bibinfo{number}{3} (\bibinfo{year}{2006}),
  \bibinfo{pages}{862--871}.
\newblock


\bibitem[Gschwindt et~al\mbox{.}(2019)]%
        {gschwindt2019can}
\bibfield{author}{\bibinfo{person}{Mirko Gschwindt}, \bibinfo{person}{Efe
  Camci}, \bibinfo{person}{Rogerio Bonatti}, \bibinfo{person}{Wenshan Wang},
  \bibinfo{person}{Erdal Kayacan}, {and} \bibinfo{person}{Sebastian Scherer}.}
  \bibinfo{year}{2019}\natexlab{}.
\newblock \showarticletitle{Can a robot become a movie director? learning
  artistic principles for aerial cinematography}. In
  \bibinfo{booktitle}{\emph{IEEE/RSJ International Conference on Intelligent
  Robots and Systems (IROS)}}. IEEE, \bibinfo{pages}{1107--1114}.
\newblock


\bibitem[He et~al\mbox{.}(2020)]%
        {he2020momentum}
\bibfield{author}{\bibinfo{person}{Kaiming He}, \bibinfo{person}{Haoqi Fan},
  \bibinfo{person}{Yuxin Wu}, \bibinfo{person}{Saining Xie}, {and}
  \bibinfo{person}{Ross Girshick}.} \bibinfo{year}{2020}\natexlab{}.
\newblock \showarticletitle{Momentum contrast for unsupervised visual
  representation learning}. In \bibinfo{booktitle}{\emph{Proceedings of the
  IEEE/CVF Conference on Computer Vision and Pattern Recognition}}.
  \bibinfo{pages}{9729--9738}.
\newblock


\bibitem[He et~al\mbox{.}(1996)]%
        {he1996virtual}
\bibfield{author}{\bibinfo{person}{Li-wei He}, \bibinfo{person}{Michael~F
  Cohen}, {and} \bibinfo{person}{David~H Salesin}.}
  \bibinfo{year}{1996}\natexlab{}.
\newblock \showarticletitle{The virtual cinematographer: A paradigm for
  automatic real-time camera control and directing}. In
  \bibinfo{booktitle}{\emph{Proceedings of the annual conference on Computer
  graphics and interactive techniques}}. \bibinfo{pages}{217--224}.
\newblock


\bibitem[He et~al\mbox{.}(2022)]%
        {he2022latent}
\bibfield{author}{\bibinfo{person}{Yingqing He}, \bibinfo{person}{Tianyu Yang},
  \bibinfo{person}{Yong Zhang}, \bibinfo{person}{Ying Shan}, {and}
  \bibinfo{person}{Qifeng Chen}.} \bibinfo{year}{2022}\natexlab{}.
\newblock \showarticletitle{Latent Video Diffusion Models for High-Fidelity
  Video Generation with Arbitrary Lengths}.
\newblock \bibinfo{journal}{\emph{arXiv preprint arXiv:2211.13221}}
  (\bibinfo{year}{2022}).
\newblock


\bibitem[Hong et~al\mbox{.}(2022)]%
        {hong2022avatarclip}
\bibfield{author}{\bibinfo{person}{Fangzhou Hong}, \bibinfo{person}{Mingyuan
  Zhang}, \bibinfo{person}{Liang Pan}, \bibinfo{person}{Zhongang Cai},
  \bibinfo{person}{Lei Yang}, {and} \bibinfo{person}{Ziwei Liu}.}
  \bibinfo{year}{2022}\natexlab{}.
\newblock \showarticletitle{AvatarCLIP: Zero-Shot Text-Driven Generation and
  Animation of 3D Avatars}.
\newblock \bibinfo{journal}{\emph{ACM Transactions on Graphics (TOG)}}
  \bibinfo{volume}{41}, \bibinfo{number}{4} (\bibinfo{year}{2022}),
  \bibinfo{pages}{1--19}.
\newblock


\bibitem[Huang et~al\mbox{.}(2021)]%
        {huang2021one}
\bibfield{author}{\bibinfo{person}{Chong Huang}, \bibinfo{person}{Yuanjie
  Dang}, \bibinfo{person}{Peng Chen}, \bibinfo{person}{Xin Yang}, {and}
  \bibinfo{person}{Kwang-Ting~Tim Cheng}.} \bibinfo{year}{2021}\natexlab{}.
\newblock \showarticletitle{One-Shot Imitation Drone Filming of Human Motion
  Videos}.
\newblock \bibinfo{journal}{\emph{IEEE Transactions on Pattern Analysis and
  Machine Intelligence}} (\bibinfo{year}{2021}).
\newblock


\bibitem[Huang et~al\mbox{.}(2019)]%
        {huang2019cvpr}
\bibfield{author}{\bibinfo{person}{Chong Huang}, \bibinfo{person}{Chuan-En
  Lin}, \bibinfo{person}{Zhenyu Yang}, \bibinfo{person}{Yan Kong},
  \bibinfo{person}{Peng Chen}, \bibinfo{person}{Xin Yang}, {and}
  \bibinfo{person}{Kwang-Ting Cheng}.} \bibinfo{year}{2019}\natexlab{}.
\newblock \showarticletitle{Learning to film from professional human motion
  videos}. In \bibinfo{booktitle}{\emph{Proceedings of the IEEE/CVF Conference
  on Computer Vision and Pattern Recognition}}. \bibinfo{pages}{4244--4253}.
\newblock


\bibitem[Huang et~al\mbox{.}(2016)]%
        {huang2016trip}
\bibfield{author}{\bibinfo{person}{Hui Huang}, \bibinfo{person}{Dani
  Lischinski}, \bibinfo{person}{Zhuming Hao}, \bibinfo{person}{Minglun Gong},
  \bibinfo{person}{Marc Christie}, {and} \bibinfo{person}{Daniel Cohen-Or}.}
  \bibinfo{year}{2016}\natexlab{}.
\newblock \showarticletitle{Trip Synopsis: 60km in 60sec}. In
  \bibinfo{booktitle}{\emph{Computer Graphics Forum}},
  Vol.~\bibinfo{volume}{35}. Wiley Online Library, \bibinfo{pages}{107--116}.
\newblock


\bibitem[Jiang et~al\mbox{.}(2021a)]%
        {jiang2021example}
\bibfield{author}{\bibinfo{person}{Hongda Jiang}, \bibinfo{person}{Marc
  Christie}, \bibinfo{person}{Xi Wang}, \bibinfo{person}{Bin Wang}, {and}
  \bibinfo{person}{Baoquan Chen}.} \bibinfo{year}{2021}\natexlab{a}.
\newblock \showarticletitle{Camera Keyframing with Style and Control}.
\newblock \bibinfo{journal}{\emph{ACM Transactions on Graphics (TOG)}}
  (\bibinfo{year}{2021}).
\newblock


\bibitem[Jiang et~al\mbox{.}(2020)]%
        {jiang2020example}
\bibfield{author}{\bibinfo{person}{Hongda Jiang}, \bibinfo{person}{Bin Wang},
  \bibinfo{person}{Xi Wang}, \bibinfo{person}{Marc Christie}, {and}
  \bibinfo{person}{Baoquan Chen}.} \bibinfo{year}{2020}\natexlab{}.
\newblock \showarticletitle{Example-driven virtual cinematography by learning
  camera behaviors}.
\newblock \bibinfo{journal}{\emph{ACM Transactions on Graphics (TOG)}}
  \bibinfo{volume}{39}, \bibinfo{number}{4} (\bibinfo{year}{2020}),
  \bibinfo{pages}{45--1}.
\newblock


\bibitem[Jiang et~al\mbox{.}(2021b)]%
        {jiang2021jointly}
\bibfield{author}{\bibinfo{person}{Xuekun Jiang}, \bibinfo{person}{Libiao Jin},
  \bibinfo{person}{Anyi Rao}, \bibinfo{person}{Linning Xu}, {and}
  \bibinfo{person}{Dahua Lin}.} \bibinfo{year}{2021}\natexlab{b}.
\newblock \showarticletitle{Jointly Learning the Attributes and Composition of
  Shots for Boundary Detection in Videos}.
\newblock \bibinfo{journal}{\emph{IEEE Transactions on Multimedia}}
  (\bibinfo{year}{2021}).
\newblock


\bibitem[Leake et~al\mbox{.}(2017)]%
        {leake2017computational}
\bibfield{author}{\bibinfo{person}{Mackenzie Leake}, \bibinfo{person}{Abe
  Davis}, \bibinfo{person}{Anh Truong}, {and} \bibinfo{person}{Maneesh
  Agrawala}.} \bibinfo{year}{2017}\natexlab{}.
\newblock \showarticletitle{Computational video editing for dialogue-driven
  scenes.}
\newblock \bibinfo{journal}{\emph{ACM Trans. Graph.}} \bibinfo{volume}{36},
  \bibinfo{number}{4} (\bibinfo{year}{2017}), \bibinfo{pages}{130--1}.
\newblock


\bibitem[Liao et~al\mbox{.}(2020)]%
        {liao2020occlusion}
\bibfield{author}{\bibinfo{person}{Junhua Liao}, \bibinfo{person}{Haihan Duan},
  \bibinfo{person}{Xin Li}, \bibinfo{person}{Haoran Xu},
  \bibinfo{person}{Yanbing Yang}, \bibinfo{person}{Wei Cai},
  \bibinfo{person}{Yanru Chen}, {and} \bibinfo{person}{Liangyin Chen}.}
  \bibinfo{year}{2020}\natexlab{}.
\newblock \showarticletitle{Occlusion Detection for Automatic Video Editing}.
  In \bibinfo{booktitle}{\emph{Proceedings of the ACM International Conference
  on Multimedia}}. \bibinfo{pages}{2255--2263}.
\newblock


\bibitem[Lino and Christie(2015)]%
        {lino2015intuitive}
\bibfield{author}{\bibinfo{person}{Christophe Lino} {and} \bibinfo{person}{Marc
  Christie}.} \bibinfo{year}{2015}\natexlab{}.
\newblock \showarticletitle{Intuitive and efficient camera control with the
  toric space}.
\newblock \bibinfo{journal}{\emph{ACM Transactions on Graphics (TOG)}}
  \bibinfo{volume}{34}, \bibinfo{number}{4} (\bibinfo{year}{2015}),
  \bibinfo{pages}{1--12}.
\newblock


\bibitem[Louarn et~al\mbox{.}(2018)]%
        {louarn2018automated}
\bibfield{author}{\bibinfo{person}{Amaury Louarn}, \bibinfo{person}{Marc
  Christie}, {and} \bibinfo{person}{Fabrice Lamarche}.}
  \bibinfo{year}{2018}\natexlab{}.
\newblock \showarticletitle{Automated staging for virtual cinematography}. In
  \bibinfo{booktitle}{\emph{Proceedings of the 11th Annual International
  Conference on Motion, Interaction, and Games}}. \bibinfo{pages}{1--10}.
\newblock


\bibitem[Louarn et~al\mbox{.}(2020)]%
        {louarn2020interactive}
\bibfield{author}{\bibinfo{person}{Amaury Louarn}, \bibinfo{person}{Quentin
  Galvane}, \bibinfo{person}{Fabrice Lamarche}, {and} \bibinfo{person}{Marc
  Christie}.} \bibinfo{year}{2020}\natexlab{}.
\newblock \showarticletitle{An interactive staging-and-shooting solver for
  virtual cinematography}.
\newblock In \bibinfo{booktitle}{\emph{Motion, Interaction and Games}}.
  \bibinfo{pages}{1--6}.
\newblock


\bibitem[Mirowski et~al\mbox{.}(2022)]%
        {mirowski2022co}
\bibfield{author}{\bibinfo{person}{Piotr Mirowski}, \bibinfo{person}{Kory~W
  Mathewson}, \bibinfo{person}{Jaylen Pittman}, {and} \bibinfo{person}{Richard
  Evans}.} \bibinfo{year}{2022}\natexlab{}.
\newblock \showarticletitle{Co-writing screenplays and theatre scripts with
  language models: An evaluation by industry professionals}.
\newblock \bibinfo{journal}{\emph{arXiv preprint arXiv:2209.14958}}
  (\bibinfo{year}{2022}).
\newblock


\bibitem[Mohanta et~al\mbox{.}(2013)]%
        {mohanta2013novel}
\bibfield{author}{\bibinfo{person}{Partha~Pratim Mohanta},
  \bibinfo{person}{Sanjoy~Kumar Saha}, {and} \bibinfo{person}{Bhabatosh
  Chanda}.} \bibinfo{year}{2013}\natexlab{}.
\newblock \showarticletitle{A novel technique for size constrained video
  storyboard generation using statistical run test and spanning tree}.
\newblock \bibinfo{journal}{\emph{International Journal of Image and Graphics}}
  \bibinfo{volume}{13}, \bibinfo{number}{01} (\bibinfo{year}{2013}),
  \bibinfo{pages}{1350001}.
\newblock


\bibitem[Moorthy et~al\mbox{.}(2020)]%
        {moorthy2020gazed}
\bibfield{author}{\bibinfo{person}{KL~Bhanu Moorthy}, \bibinfo{person}{Moneish
  Kumar}, \bibinfo{person}{Ramanathan Subramanian}, {and}
  \bibinfo{person}{Vineet Gandhi}.} \bibinfo{year}{2020}\natexlab{}.
\newblock \showarticletitle{Gazed--gaze-guided cinematic editing of wide-angle
  monocular video recordings}. In \bibinfo{booktitle}{\emph{Proceedings of the
  CHI Conference on Human Factors in Computing Systems}}.
  \bibinfo{pages}{1--11}.
\newblock


\bibitem[Nvidia(2023)]%
        {omniverse}
\bibfield{author}{\bibinfo{person}{Nvidia}.} \bibinfo{year}{2023}\natexlab{}.
\newblock \bibinfo{title}{Omniverse Platform}.
\newblock \bibinfo{howpublished}{\url{https://www.nvidia.com/omniverse/}}.
\newblock


\bibitem[Oskam et~al\mbox{.}(2009)]%
        {oskam2009visibility}
\bibfield{author}{\bibinfo{person}{Thomas Oskam}, \bibinfo{person}{Robert~W
  Sumner}, \bibinfo{person}{Nils Thuerey}, {and} \bibinfo{person}{Markus
  Gross}.} \bibinfo{year}{2009}\natexlab{}.
\newblock \showarticletitle{Visibility transition planning for dynamic camera
  control}. In \bibinfo{booktitle}{\emph{Proceedings of the ACM
  SIGGRAPH/Eurographics Symposium on Computer Animation}}.
  \bibinfo{pages}{55--65}.
\newblock


\bibitem[Pan et~al\mbox{.}(2021)]%
        {pan2021videomoco}
\bibfield{author}{\bibinfo{person}{Tian Pan}, \bibinfo{person}{Yibing Song},
  \bibinfo{person}{Tianyu Yang}, \bibinfo{person}{Wenhao Jiang}, {and}
  \bibinfo{person}{Wei Liu}.} \bibinfo{year}{2021}\natexlab{}.
\newblock \showarticletitle{Videomoco: Contrastive video representation
  learning with temporally adversarial examples}. In
  \bibinfo{booktitle}{\emph{Proceedings of the IEEE/CVF Conference on Computer
  Vision and Pattern Recognition}}. \bibinfo{pages}{11205--11214}.
\newblock


\bibitem[Pardo et~al\mbox{.}(2021a)]%
        {pardo2021learning}
\bibfield{author}{\bibinfo{person}{Alejandro Pardo}, \bibinfo{person}{Fabian
  Caba}, \bibinfo{person}{Juan~Le{\'o}n Alc{\'a}zar}, \bibinfo{person}{Ali~K
  Thabet}, {and} \bibinfo{person}{Bernard Ghanem}.}
  \bibinfo{year}{2021}\natexlab{a}.
\newblock \showarticletitle{Learning to Cut by Watching Movies}. In
  \bibinfo{booktitle}{\emph{Proceedings of the IEEE/CVF International
  Conference on Computer Vision}}. \bibinfo{pages}{6858--6868}.
\newblock


\bibitem[Pardo et~al\mbox{.}(2021b)]%
        {pardo2021moviecuts}
\bibfield{author}{\bibinfo{person}{Alejandro Pardo},
  \bibinfo{person}{Fabian~Caba Heilbron}, \bibinfo{person}{Juan~Le{\'o}n
  Alc{\'a}zar}, \bibinfo{person}{Ali Thabet}, {and} \bibinfo{person}{Bernard
  Ghanem}.} \bibinfo{year}{2021}\natexlab{b}.
\newblock \showarticletitle{MovieCuts: A New Dataset and Benchmark for Cut Type
  Recognition}.
\newblock \bibinfo{journal}{\emph{arXiv preprint arXiv:2109.05569}}
  (\bibinfo{year}{2021}).
\newblock


\bibitem[Pizzi et~al\mbox{.}(2010)]%
        {pizzi2010automatic}
\bibfield{author}{\bibinfo{person}{David Pizzi}, \bibinfo{person}{Jean-Luc
  Lugrin}, \bibinfo{person}{Alex Whittaker}, {and} \bibinfo{person}{Marc
  Cavazza}.} \bibinfo{year}{2010}\natexlab{}.
\newblock \showarticletitle{Automatic generation of game level solutions as
  storyboards}.
\newblock \bibinfo{journal}{\emph{IEEE Transactions on Computational
  Intelligence and AI in Games}} \bibinfo{volume}{2}, \bibinfo{number}{3}
  (\bibinfo{year}{2010}), \bibinfo{pages}{149--161}.
\newblock


\bibitem[Rao et~al\mbox{.}(2022a)]%
        {rao2022temporal}
\bibfield{author}{\bibinfo{person}{Anyi Rao}, \bibinfo{person}{Xuekun Jiang},
  \bibinfo{person}{Sichen Wang}, \bibinfo{person}{Yuwei Guo},
  \bibinfo{person}{Zihao Liu}, \bibinfo{person}{Bo Dai}, \bibinfo{person}{Long
  Pang}, \bibinfo{person}{Xiaoyu Wu}, \bibinfo{person}{Dahua Lin}, {and}
  \bibinfo{person}{Libiao Jin}.} \bibinfo{year}{2022}\natexlab{a}.
\newblock \showarticletitle{Temporal and Contextual Transformer for
  Multi-Camera Editing of TV Shows}.
\newblock \bibinfo{journal}{\emph{arXiv preprint arXiv:2210.08737}}
  (\bibinfo{year}{2022}).
\newblock


\bibitem[Rao et~al\mbox{.}(2020)]%
        {rao2020unified}
\bibfield{author}{\bibinfo{person}{Anyi Rao}, \bibinfo{person}{Jiaze Wang},
  \bibinfo{person}{Linning Xu}, \bibinfo{person}{Xuekun Jiang},
  \bibinfo{person}{Qingqiu Huang}, \bibinfo{person}{Bolei Zhou}, {and}
  \bibinfo{person}{Dahua Lin}.} \bibinfo{year}{2020}\natexlab{}.
\newblock \showarticletitle{A Unified Framework for Shot Type Classification
  Based on Subject Centric Lens}. In \bibinfo{booktitle}{\emph{The European
  Conference on Computer Vision (ECCV)}}.
\newblock


\bibitem[Rao et~al\mbox{.}(2022b)]%
        {rao2022shoot360}
\bibfield{author}{\bibinfo{person}{Anyi Rao}, \bibinfo{person}{Linning Xu},
  {and} \bibinfo{person}{Dahua Lin}.} \bibinfo{year}{2022}\natexlab{b}.
\newblock \showarticletitle{Shoot360: Normal View Video Creation from City
  Panorama Footage}. In \bibinfo{booktitle}{\emph{ACM SIGGRAPH 2022 Conference
  Proceedings}}. \bibinfo{pages}{1--9}.
\newblock


\bibitem[Richter et~al\mbox{.}(2016)]%
        {richter2016playing}
\bibfield{author}{\bibinfo{person}{Stephan~R Richter}, \bibinfo{person}{Vibhav
  Vineet}, \bibinfo{person}{Stefan Roth}, {and} \bibinfo{person}{Vladlen
  Koltun}.} \bibinfo{year}{2016}\natexlab{}.
\newblock \showarticletitle{Playing for data: Ground truth from computer
  games}. In \bibinfo{booktitle}{\emph{European conference on computer
  vision}}. Springer, \bibinfo{pages}{102--118}.
\newblock


\bibitem[Ronfard et~al\mbox{.}(2022)]%
        {ronfard2022prose}
\bibfield{author}{\bibinfo{person}{Rmi Ronfard}, \bibinfo{person}{Vineet
  Gandhi}, \bibinfo{person}{Laurent Boiron}, {and} \bibinfo{person}{A
  Murukutla}.} \bibinfo{year}{2022}\natexlab{}.
\newblock \showarticletitle{The prose storyboard language: A tool for
  annotating and directing movies (version 2.0, revised and illustrated
  edition)}. In \bibinfo{booktitle}{\emph{Eurographics Workshop on Intelligent
  Cinematography and Editing}}, Vol.~\bibinfo{volume}{4}.
\newblock


\bibitem[Sanokho et~al\mbox{.}(2014)]%
        {sanokho2014camera}
\bibfield{author}{\bibinfo{person}{Cunka~Bassirou Sanokho},
  \bibinfo{person}{Clement Desoche}, \bibinfo{person}{Billal Merabti},
  \bibinfo{person}{Tsai-Yen Li}, {and} \bibinfo{person}{Marc Christie}.}
  \bibinfo{year}{2014}\natexlab{}.
\newblock \showarticletitle{Camera Motion Graphs.}. In
  \bibinfo{booktitle}{\emph{Symposium on Computer Animation}}. Citeseer,
  \bibinfo{pages}{177--188}.
\newblock


\bibitem[Savva et~al\mbox{.}(2019)]%
        {savva2019habitat}
\bibfield{author}{\bibinfo{person}{Manolis Savva}, \bibinfo{person}{Abhishek
  Kadian}, \bibinfo{person}{Oleksandr Maksymets}, \bibinfo{person}{Yili Zhao},
  \bibinfo{person}{Erik Wijmans}, \bibinfo{person}{Bhavana Jain},
  \bibinfo{person}{Julian Straub}, \bibinfo{person}{Jia Liu},
  \bibinfo{person}{Vladlen Koltun}, \bibinfo{person}{Jitendra Malik},
  {et~al\mbox{.}}} \bibinfo{year}{2019}\natexlab{}.
\newblock \showarticletitle{Habitat: A platform for embodied ai research}. In
  \bibinfo{booktitle}{\emph{Proceedings of the IEEE/CVF International
  Conference on Computer Vision}}. \bibinfo{pages}{9339--9347}.
\newblock


\bibitem[Shah et~al\mbox{.}(2018)]%
        {shah2018airsim}
\bibfield{author}{\bibinfo{person}{Shital Shah}, \bibinfo{person}{Debadeepta
  Dey}, \bibinfo{person}{Chris Lovett}, {and} \bibinfo{person}{Ashish Kapoor}.}
  \bibinfo{year}{2018}\natexlab{}.
\newblock \showarticletitle{Airsim: High-fidelity visual and physical
  simulation for autonomous vehicles}. In \bibinfo{booktitle}{\emph{Field and
  service robotics}}. Springer, \bibinfo{pages}{621--635}.
\newblock


\bibitem[Shridhar et~al\mbox{.}(2020)]%
        {shridhar2020alfred}
\bibfield{author}{\bibinfo{person}{Mohit Shridhar}, \bibinfo{person}{Jesse
  Thomason}, \bibinfo{person}{Daniel Gordon}, \bibinfo{person}{Yonatan Bisk},
  \bibinfo{person}{Winson Han}, \bibinfo{person}{Roozbeh Mottaghi},
  \bibinfo{person}{Luke Zettlemoyer}, {and} \bibinfo{person}{Dieter Fox}.}
  \bibinfo{year}{2020}\natexlab{}.
\newblock \showarticletitle{Alfred: A benchmark for interpreting grounded
  instructions for everyday tasks}. In \bibinfo{booktitle}{\emph{Proceedings of
  the IEEE/CVF conference on computer vision and pattern recognition}}.
  \bibinfo{pages}{10740--10749}.
\newblock


\bibitem[Singer et~al\mbox{.}(2022)]%
        {singer2022make}
\bibfield{author}{\bibinfo{person}{Uriel Singer}, \bibinfo{person}{Adam
  Polyak}, \bibinfo{person}{Thomas Hayes}, \bibinfo{person}{Xi Yin},
  \bibinfo{person}{Jie An}, \bibinfo{person}{Songyang Zhang},
  \bibinfo{person}{Qiyuan Hu}, \bibinfo{person}{Harry Yang},
  \bibinfo{person}{Oron Ashual}, \bibinfo{person}{Oran Gafni}, {et~al\mbox{.}}}
  \bibinfo{year}{2022}\natexlab{}.
\newblock \showarticletitle{Make-a-video: Text-to-video generation without
  text-video data}.
\newblock \bibinfo{journal}{\emph{arXiv preprint arXiv:2209.14792}}
  (\bibinfo{year}{2022}).
\newblock


\bibitem[Starke et~al\mbox{.}(2019)]%
        {starke2019neural}
\bibfield{author}{\bibinfo{person}{Sebastian Starke}, \bibinfo{person}{He
  Zhang}, \bibinfo{person}{Taku Komura}, {and} \bibinfo{person}{Jun Saito}.}
  \bibinfo{year}{2019}\natexlab{}.
\newblock \showarticletitle{Neural state machine for character-scene
  interactions.}
\newblock \bibinfo{journal}{\emph{ACM Trans. Graph.}} \bibinfo{volume}{38},
  \bibinfo{number}{6} (\bibinfo{year}{2019}), \bibinfo{pages}{209--1}.
\newblock


\bibitem[Truong and Agrawala(2019)]%
        {truong2019tool}
\bibfield{author}{\bibinfo{person}{Anh Truong} {and} \bibinfo{person}{Maneesh
  Agrawala}.} \bibinfo{year}{2019}\natexlab{}.
\newblock \showarticletitle{A Tool for Navigating and Editing 360 Video of
  Social Conversations into Shareable Highlights.}. In
  \bibinfo{booktitle}{\emph{Graphics Interface}}. \bibinfo{pages}{14--1}.
\newblock


\bibitem[Truong et~al\mbox{.}(2018)]%
        {truong2018extracting}
\bibfield{author}{\bibinfo{person}{Anh Truong}, \bibinfo{person}{Sara Chen},
  \bibinfo{person}{Ersin Yumer}, \bibinfo{person}{David Salesin}, {and}
  \bibinfo{person}{Wilmot Li}.} \bibinfo{year}{2018}\natexlab{}.
\newblock \showarticletitle{Extracting regular fov shots from 360 event
  footage}. In \bibinfo{booktitle}{\emph{Proceedings of the CHI Conference on
  Human Factors in Computing Systems}}. \bibinfo{pages}{1--11}.
\newblock


\bibitem[Unity(2023)]%
        {unity}
\bibfield{author}{\bibinfo{person}{Unity}.} \bibinfo{year}{2023}\natexlab{}.
\newblock \bibinfo{title}{Real-Time Development Platform}.
\newblock \bibinfo{howpublished}{\url{https://unity.com/}}.
\newblock


\bibitem[Unreal(2023)]%
        {unreal}
\bibfield{author}{\bibinfo{person}{Unreal}.} \bibinfo{year}{2023}\natexlab{}.
\newblock \bibinfo{title}{Real-Time 3D Creation Tool}.
\newblock \bibinfo{howpublished}{\url{https://www.unrealengine.com/}}.
\newblock


\bibitem[Wang et~al\mbox{.}(2021)]%
        {wang2021scene}
\bibfield{author}{\bibinfo{person}{Jingbo Wang}, \bibinfo{person}{Sijie Yan},
  \bibinfo{person}{Bo Dai}, {and} \bibinfo{person}{Dahua Lin}.}
  \bibinfo{year}{2021}\natexlab{}.
\newblock \showarticletitle{Scene-aware Generative Network for Human Motion
  Synthesis}. In \bibinfo{booktitle}{\emph{Proceedings of the IEEE/CVF
  Conference on Computer Vision and Pattern Recognition}}.
  \bibinfo{pages}{12206--12215}.
\newblock


\bibitem[Wang et~al\mbox{.}(2018)]%
        {wang2018temporal}
\bibfield{author}{\bibinfo{person}{Limin Wang}, \bibinfo{person}{Yuanjun
  Xiong}, \bibinfo{person}{Zhe Wang}, \bibinfo{person}{Yu Qiao},
  \bibinfo{person}{Dahua Lin}, \bibinfo{person}{Xiaoou Tang}, {and}
  \bibinfo{person}{Luc Van~Gool}.} \bibinfo{year}{2018}\natexlab{}.
\newblock \showarticletitle{Temporal segment networks for action recognition in
  videos}.
\newblock \bibinfo{journal}{\emph{IEEE transactions on pattern analysis and
  machine intelligence}} \bibinfo{volume}{41}, \bibinfo{number}{11}
  (\bibinfo{year}{2018}), \bibinfo{pages}{2740--2755}.
\newblock


\bibitem[Wang et~al\mbox{.}(2019)]%
        {wang2019write}
\bibfield{author}{\bibinfo{person}{Miao Wang}, \bibinfo{person}{Guo-Wei Yang},
  \bibinfo{person}{Shi-Min Hu}, \bibinfo{person}{Shing-Tung Yau}, {and}
  \bibinfo{person}{Ariel Shamir}.} \bibinfo{year}{2019}\natexlab{}.
\newblock \showarticletitle{Write-a-video: computational video montage from
  themed text.}
\newblock \bibinfo{journal}{\emph{ACM Trans. Graph.}} \bibinfo{volume}{38},
  \bibinfo{number}{6} (\bibinfo{year}{2019}), \bibinfo{pages}{177--1}.
\newblock


\bibitem[Wikipedia(2023)]%
        {storyboard}
\bibfield{author}{\bibinfo{person}{Wikipedia}.}
  \bibinfo{year}{2023}\natexlab{}.
\newblock \bibinfo{title}{Storyboard}.
\newblock
  \bibinfo{howpublished}{\url{https://en.wikipedia.org/wiki/Storyboard}}.
\newblock


\bibitem[Wu et~al\mbox{.}(2018)]%
        {wu2018thinking}
\bibfield{author}{\bibinfo{person}{Hui-Yin Wu}, \bibinfo{person}{Francesca
  Pal{\`u}}, \bibinfo{person}{Roberto Ranon}, {and} \bibinfo{person}{Marc
  Christie}.} \bibinfo{year}{2018}\natexlab{}.
\newblock \showarticletitle{Thinking like a director: Film editing patterns for
  virtual cinematographic storytelling}.
\newblock \bibinfo{journal}{\emph{ACM Transactions on Multimedia Computing,
  Communications, and Applications (TOMM)}} \bibinfo{volume}{14},
  \bibinfo{number}{4} (\bibinfo{year}{2018}), \bibinfo{pages}{1--22}.
\newblock


\bibitem[Ye and Baldwin(2008)]%
        {ye2008towards}
\bibfield{author}{\bibinfo{person}{Patrick Ye} {and} \bibinfo{person}{Timothy
  Baldwin}.} \bibinfo{year}{2008}\natexlab{}.
\newblock \showarticletitle{Towards Automatic Animated Storyboarding.}. In
  \bibinfo{booktitle}{\emph{AAAI}}. \bibinfo{pages}{578--583}.
\newblock


\bibitem[Yoo et~al\mbox{.}(2021)]%
        {yoo2021virtual}
\bibfield{author}{\bibinfo{person}{Jung~Eun Yoo}, \bibinfo{person}{Kwanggyoon
  Seo}, \bibinfo{person}{Sanghun Park}, \bibinfo{person}{Jaedong Kim},
  \bibinfo{person}{Dawon Lee}, {and} \bibinfo{person}{Junyong Noh}.}
  \bibinfo{year}{2021}\natexlab{}.
\newblock \showarticletitle{Virtual Camera Layout Generation using a Reference
  Video}. In \bibinfo{booktitle}{\emph{Proceedings of the CHI Conference on
  Human Factors in Computing Systems}}. \bibinfo{pages}{1--11}.
\newblock


\bibitem[Zhang et~al\mbox{.}(2018)]%
        {zhang2018mode}
\bibfield{author}{\bibinfo{person}{He Zhang}, \bibinfo{person}{Sebastian
  Starke}, \bibinfo{person}{Taku Komura}, {and} \bibinfo{person}{Jun Saito}.}
  \bibinfo{year}{2018}\natexlab{}.
\newblock \showarticletitle{Mode-adaptive neural networks for quadruped motion
  control}.
\newblock \bibinfo{journal}{\emph{ACM Transactions on Graphics (TOG)}}
  \bibinfo{volume}{37}, \bibinfo{number}{4} (\bibinfo{year}{2018}),
  \bibinfo{pages}{1--11}.
\newblock


\bibitem[Zhong et~al\mbox{.}(2021)]%
        {zhong2021aesthetic}
\bibfield{author}{\bibinfo{person}{Lei Zhong}, \bibinfo{person}{Feng-Heng Li},
  \bibinfo{person}{Hao-Zhi Huang}, \bibinfo{person}{Yong Zhang},
  \bibinfo{person}{Shao-Ping Lu}, {and} \bibinfo{person}{Jue Wang}.}
  \bibinfo{year}{2021}\natexlab{}.
\newblock \showarticletitle{Aesthetic-guided outward image cropping}.
\newblock \bibinfo{journal}{\emph{ACM Transactions on Graphics (TOG)}}
  \bibinfo{volume}{40}, \bibinfo{number}{6} (\bibinfo{year}{2021}),
  \bibinfo{pages}{1--13}.
\newblock


\end{thebibliography}
